\documentclass[journal]{IEEEtran}
\usepackage{amsmath,amssymb,graphicx,psfrag}
\usepackage{hyperref,mathrsfs,subfigure,epstopdf}
\usepackage{graphics}
\usepackage{epsfig}
\usepackage[usenames,dvipsnames]{color}
\usepackage{comment}

\newtheorem{proposition}{\bf Proposition}

\newtheorem{property}{\bf Property}

\newtheorem{definition}{Definition}
\newtheorem{lemma}{Lemma}

\newcommand{\EE}{\mathbb{E}} 
\newcommand{\RR}{\mathbb{R}} 

\newcommand{\ee}{{\rm e}}
\newcommand{\dd}{{\rm\,d}} 

\newcommand{\av}{{\bf a}}
\newcommand{\bv}{{\bf b}}

\newcommand{\ev}{{\bf e}}

\newcommand{\pv}{{\bf p}}
\newcommand{\qv}{{\bf q}}

\newcommand{\xv}{{\bf x}}
\newcommand{\yv}{{\bf y}}

\newcommand{\zerov}{{\bf 0}}

\newcommand{\Am}{{\bf A}}
\newcommand{\Bm}{{\bf B}}
\newcommand{\Cm}{{\bf C}}
\newcommand{\Dm}{{\bf D}}
\newcommand{\Em}{{\bf E}}
\newcommand{\Fm}{{\bf F}}
\newcommand{\Gm}{{\bf G}}
\newcommand{\Hm}{{\bf H}}
\newcommand{\Id}{{\bf I}}

\newcommand{\Lm}{{\bf L}}
\newcommand{\Mm}{{\bf M}}
\newcommand{\Nm}{{\bf N}}

\newcommand{\Pm}{{\bf P}}

\newcommand{\Rm}{{\bf R}}
\newcommand{\Sm}{{\bf S}}
\newcommand{\Tm}{{\bf T}}
\newcommand{\Um}{{\bf U}}
\newcommand{\Wm}{{\bf W}}
\newcommand{\Vm}{{\bf V}}
\newcommand{\Xm}{{\bf X}}
\newcommand{\Ym}{{\bf Y}}
\newcommand{\Zm}{{\bf Z}}

\newcommand{\Bc}{{\cal B}}

\newcommand{\Dc}{{\cal D}}

\newcommand{\Fc}{{\cal F}}

\newcommand{\Ic}{{\cal I}}

\newcommand{\Sc}{{\cal S}}
  
\newcommand{\Uc}{{\cal U}}
\newcommand{\Wc}{{\cal W}}
\newcommand{\Vc}{{\cal V}}
\newcommand{\Xc}{{\cal X}}

\newcommand{\psiv}{\boldsymbol{\psi}}

\newcommand{\Lambdam}{\boldsymbol{\Lambda}}
\newcommand{\Deltam}{\boldsymbol{\Delta}}
\newcommand{\Sigmam}{\boldsymbol{\Sigma}}
\newcommand{\Phim}{\boldsymbol{\Phi}}

\newcommand{\Psim}{\boldsymbol{\Psi}}
\newcommand{\Thetam}{\boldsymbol{\Theta}}
\newcommand{\Omegam}{\boldsymbol{\Omega}}
\newcommand{\Xim}{\boldsymbol{\Xi}}

\newcommand{\diag}{{\hbox{diag}}}

\def\trace{\mathsf{Tr}}

\def\Herm{\mathsf{^H}}
\def\Tran{\mathsf{^T}}

\newcommand{\plus}{\mathord{+}}
\newcommand{\minus}{\mathord{-}}
\def\ky{K(\Ym)}

\def\non{\nonumber\\}

\title{Closed-form Output Statistics of MIMO Block-Fading Channels}
\author{
Giuseppa Alfano,~\IEEEmembership{Member,~IEEE,}
Carla-Fabiana Chiasserini,~\IEEEmembership{Senior Member,~IEEE,}  \\
Alessandro Nordio,~\IEEEmembership{Member,~IEEE,} 
Siyuan Zhou
\thanks{Alessandro~Nordio is with IEIIT-CNR (Institute of Electronics,
  Telecommunications and Information Engineering of the National Research Council of Italy), 
  Italy, email: alessandro.nordio@ieiit.cnr.it.}%
\thanks{Giuseppa Alfano, Siyuan Zhou, and C.-F. Chiasserini are with the Dipartimento di
  Elettronica and Telecomunicazioni, Politecnico di Torino, Torino, Italy, email:
  alfano@tlc.polito.it, siyuan.zhou@polito.it, chiasserini@polito.it.}%
}
\begin{document}
\maketitle

\begin{abstract}
The information that can be transmitted through a wireless channel, with multiple-antenna equipped
transmitter and receiver, is crucially influenced by the channel behavior as well as by the
structure of the input signal.  We characterize in closed form the probability density function
(pdf) of the output of MIMO block-fading channels, for an arbitrary SNR value.  Our results provide
compact expressions for such output statistics, paving the way to a more detailed analytical
information-theoretic exploration of communications in presence of block fading. The analysis is
carried out assuming two different structures for the input signal: the i.i.d. Gaussian distribution
and a product form that has been proved to be optimal for non-coherent communication, i.e., in
absence of any channel state information.  When the channel is fed by an i.i.d. Gaussian input, we
assume the Gramian of the channel matrix to be unitarily invariant and derive the output statistics
in both the noise-limited and the interference-limited scenario, considering different fading
distributions. When the product-form input is adopted, we provide the expressions of the output pdf
as the relationship between the overall number of antennas and the fading coherence length varies.
We also highlight the relation between our newly derived expressions and the results already
available in the literature, and, for some cases, we numerically compute the mutual information,
based on the proposed expression of the output statistics.

\end{abstract}


\section{Introduction}

The availability of an explicit statistical characterization of the output of a wireless channel,
impaired by additive and multiplicative random disturbance, is of paramount importance to
communication- and information-theoretic purposes. Indeed, a closed-form expression for the output
probability density function (pdf) is relevant for the evaluation of the ergodic mutual information
between the input and the output signals of a randomly faded channel~\cite{Alfano2004}.
It also turns out to be crucial in the finite block-length regime, in order to
characterize the information density of the communication at hand~\cite{Princeton}.

In spite of its importance, few explicit results are available in the literature for the output
signal pdf in the case of MIMO block-independent fading channels. The works
in~\cite{LozanoNihar,closedform,durisi} all focus on the case of block-Rayleigh fading.  In these
papers, the output statistics are derived under different assumptions on the relative values of the
number of involved antennas and of the coherence length of the fading.  The input distribution, too,
plays a crucial role in the cited derivations.  More specifically, in \cite{LozanoNihar} the authors
assume the input to be i.i.d. Gaussian and investigate the behavior of the output distribution as
the fading coherence length varies from being quite short to very long, compared to the overall
number of transmit and receive antennas. In both \cite{closedform} and \cite{durisi}, instead, the
input is assumed to be given by the product of a diagonal matrix (representing the power allocation
over the transmit antennas) times an isotropically distributed matrix with unitary columns.  The
main difference between the two papers is in the assumption on the fading duration. Indeed, the
first one focuses on the case where the coherence length of the Rayleigh fading is greater than the
number of involved antennas; in this case, the high Signal to Noise Ratio (SNR)-optimal power
allocation matrix turns out to be a scaled version of the identity matrix \cite{tse}. The study in
\cite{durisi}, instead, solves the problem of characterizing, again in the high-SNR regime, the
optimal power allocation profile, assuming the fading coherence length to be shorter, compared to
the number of involved antennas. In the latter case, indeed, the diagonal matrix of the power
allocation is characterized by the eigenvalues of a matrix-variate Beta joint distribution of the
entries \cite{durisi}.

In this paper, we consider both the input models described above, and derive closed form expressions
of the output pdf in presence of a multiple-antenna channel affected by additive noise and
block-fading.  In particular, in the case of i.i.d. Gaussian input, our procedure allows the
derivation of a closed-form expression for the output statistics of channels with unitarily
invariant fading law. Apart from the canonical i.i.d. Rayleigh fading, already treated in
\cite{LozanoNihar}, this encompasses the Rician channel with scalar Line-of-Sight (LOS) matrix,
whose analysis was previously limited to the evaluation of the fading number \cite{Mosscalar}, and
the LOS MIMO \cite{Oien} with a certain amount of residual scattering. Also, we provide results for
the Land Mobile Satellite (LMS) with scalar average power LOS matrix \cite[Property I]{lmsmimo} and
for the above cases of MIMO Rayleigh and Rician fading communications impaired by Rayleigh-faded
co-channel interference \cite{Lozinterf}. We remark that the expressions of the output pdf that we
derive hold for any arbitrary value of SNR.

The paper is organized as follows.  Section \ref{sec:background} introduces the notations used
throughout the paper, the communication model of the wireless system and relevant mathematical
background.  Section \ref{sec:analysis1} presents the analytical derivation of the output pdfs, as
the channel is fed by an i.i.d.  Gaussian input.  Section \ref{sec:analysis2} provides the output
pdfs in presence of optimized product form in Rayleigh block-fading channels.  Finally, Section
\ref{sec:conclusion} concludes the paper.

\section{Preliminaries  and Communication Model\label{sec:background}}

\subsection{Notations}

\subsubsection{Vectors and matrices}
Throughout the paper, uppercase and lowercase boldface letters denote matrices and vectors,
respectively. The identity matrix is denoted by $\Id$. The pdf of a random matrix $\Am$,
$p_{\Am}(\Am)$, is simply indicated with $p(\Am)$, except when referring to $\Am$ is needed for
clarity.  $\EE[\cdot]$ represents statistical expectation, $(\cdot)\Herm$ indicates the conjugate
transpose operator, $\trace\{\cdot\}$ denotes the trace of a square matrix, and $\|\cdot\|$ stands
for the Euclidean norm\footnote{As applied to a matrix, we mean $\|\Am\|^2=\trace\{\Am\Herm\Am\}$.}.
Also, we indicate with $\{a_{ij}\}$ the matrix whose elements are $a_{ij}$ and with $|\Am|$, or
$|\{a_{ij}\}|$, the determinant of matrix $\Am$.  We often employ the following property of the
determinant:
\begin{property}
Let $\Fm = \{f_{ij}\}$ be an $m\times m$ matrix where $f_{ij}=\alpha
a_i b_{ij}c_j$. Then,  
\begin{equation}
  |\Fm| = \alpha^m |\{b_{ij}\}|\prod_{i=1}^m a_i \prod_{j=1}^m c_j\,.
\label{eq:determinant_property}
\end{equation}
\end{property}

\subsubsection{Complex multivariate Gamma function}
$\Gamma_m(a)$ is the complex multivariate Gamma function defined as~\cite{james}:
\[ \Gamma_m(a) = \pi_m\prod_{\ell=1}^{m} \Gamma(a -\ell+1)\]
with $m$ being a non-negative integer and 
\[ \pi_m = \pi^{m(m-1)/2}\,.\]

\subsubsection{Vandermonde determinant}
Let $\Am$ be an $m \times m$ Hermitian matrix with eigenvalues $a_1,\ldots,a_m$.
Then the Vandermonde determinant of $\Am$ is defined as~\cite[eq. (2.10)]{McKay}:
\begin{equation}
\label{eq:V_vandermonde}
\Vc(\Am) = \prod_{1 \leq i < j \leq m} (a_i - a_j ) \,,
\end{equation}
where we assume the eigenvalues to be ordered in decreasing order so that $\Vc(\Am)$ is non
negative. Moreover, for any constant $c$, we have $\Vc(c\Am) = c^{m(m-1)/2}\Vc(\Am)$.  Let also
$\Fm=\{f_i(a_j)\}$, $i,j=1,\ldots,m$, be an $m\times m$ matrix, where $f_i(\cdot)$'s are any
differentiable functions.  Clearly, if the eigenvalues of $\Am$ are not distinct, $\Vc(\Am)=0$
and $|\Fm|=0$. In such a case, the ratio $|\Fm|/\Vc(\Am)$, which appears in the density of
many matrices that we study in the following, can be evaluated by applying l'H\^opital's rule.
More precisely, let $n$ be an integer such that $0 < n < m$, then~\cite[Lemma 5]{fraclimit}
\begin{equation}
\lim_{a_{n+1},\ldots,a_m \to a}\frac{|\Fm|}{\Vc(\Am)} =
\frac{\pi_m\Gamma_n(m)}{\pi_n\Gamma_m(m)} \frac{|\widetilde{\Fm}|}{\Vc(\widetilde{\Am})}|\widetilde{\Am} -a\Id|^{n-m}
\label{eq:limit}
\end{equation}
where $\widetilde{\Am}$ is of size $n\times n$ and has eigenvalues $a_1,\ldots, a_n$ and
\[  (\widetilde{\Fm})_{ij} = \left\{ 
\begin{array}{ll}
  f_i(a_j)  & i=1,\ldots,m;\,\,j=1,\ldots,n \\
  f_i^{(m-j)}(a) & i=1,\ldots,m;\,\,j=n+1,\ldots,m 
\end{array}   
 \right. \]
with $f_i^{(k)}(\cdot)$ denoting the $k$-th derivative of $f_i(\cdot)$. For $n=0$, we have
\begin{equation} \lim_{a_{1},\ldots,a_m \to a}\frac{|\Fm|}{\Vc(\Am)} = \frac{\pi_m}{\Gamma_m(m)} |\widetilde{\Fm}| \label{eq:limit_n0}\end{equation}
where $(\widetilde{\Fm})_{ij} =f_i^{(m-j)}(a)$,  $i,j=1,\ldots,m$.

\subsubsection{Generalized hypergeometric function}
The generalized hypergeometric function is defined as $_pF_q(\av; \bv; \Xc)$, where
$\av=[a_1,\ldots,a_p]\Tran$, $\bv=[b_1,\ldots,b_q]\Tran$, and $\Xc$ is a set of arguments that can
be either scalars or square matrices~\cite{abram}. In the case of a single scalar argument,
$\Xc=\{x\}$, the generalized hypergeometric function is defined as in~\cite[eq. (2.24)]{McKay}:
\begin{equation}
\label{eq:pFq_definition}
 {_pF_q}(\av; \bv; x) = \sum_{k=0}^{\infty}\frac{[\av]_{k}}{[\bv]_{k}} \frac{x^k}{k!}
\end{equation}
where $[\av]_{k}=\prod_{i=1}^p [a_i]_{k}$, $[\bv]_{k}=\prod_{j=1}^q [b_j]_{k}$, and
$[z]_k=\Gamma(z+k)/\Gamma(z)$ denotes the Pochhammer symbol.  Note that ${_0F_0}(; ;
  x)=\ee^{x}$, and ${_1F_0}(a; ; x)=(1-x)^{-a}$. The function ${_0F_1}(;b; x)$ is closely related to the
  Bessel's function, and in the literature functions ${_1F_1}(a; b; x)$ and ${_2F_1}(a_1,a_2;
  b; x)$ are also called {\em confluent hypergeometric function of the first kind} and {\em Gauss's
    hypergeometric function}, respectively.

The generalized hypergeometric function of two matrix arguments, $\Xc=\{\Phim, \Psim\}$, both of
size $m\times m$, can be written through hypergeometric functions of scalar arguments
as~\cite[eq. (2.34)]{McKay}
\begin{equation}
 {_pF_q}(\av; \bv; \Phim,\Psim) = c\frac{|\{{_pF_q}(\tilde{\av}; \tilde{\bv}; \phi_h\psi_k)\}|}{\Vc(\Phim)\Vc(\Psim)}
\label{eq:pFq_2arguments}
\end{equation}
$h,k=1,\ldots,m$, where the constant $c$ is given by \cite{McKay} 
\[  c = \frac{\Gamma_m(m)}{\pi_m^{q-p+1}}\left[\prod_{j=1}^q\frac{\Gamma_m(b_j)}{(b_j-m)!^m}\right]\left[\prod_{i=1}^p\frac{(a_i-m)!^m}{\Gamma_m(a_i)}\right]\,, \]
$\tilde{a}_i=a_i-m+1$, $i=1,\ldots,p$, $\tilde{b}_j=b_j-m+1$, $j=1,\ldots,q$, and the eigenvalues of
$\Phim$ and $\Psim$ are denoted by $\phi_1,\ldots,\phi_m$ and $\psi_1,\ldots,\psi_m$, respectively.

The $\ell$-th derivative of the generalized hypergeometric function
${_pF_q}(\av; \bv; sx)$ is given by~\cite{abram}: 
\begin{equation}
\frac{\dd^\ell}{\dd x^\ell}{_pF_q}(\av; \bv; sx) = s^\ell\frac{(\av)_\ell}{(\bv)_\ell}{_pF_q}(\tilde{\av}; \tilde{\bv}; sx)
\label{eq:hypergeometric_derivative}
\end{equation}
where $\tilde{a}_i=a_i+\ell$, $\tilde{b}_j=b_j+\ell$, $i=1,\ldots,p$, $j=1,\ldots,q$, and $s$ is a
parameter.

\subsubsection{Matrix spaces}
We will denote by $\Uc(m)$ the unitary group of size $m$ and by $\Sc(m,n)$ the Stiefel manifold of
$m\times n$ matrices~\cite[Sec. II.C]{tse}.  The region defined by the Stiefel manifold $\Sc(m,n)$,
with $m\ge n$, is compact and has volume $|\Sc(m,n)| = 2^n \pi^{mn}/\Gamma_n(m)$.  When $m=n$, the
Stiefel manifold is a unitary group and its volume is given by $|\Uc(m)| = 2^m
\pi^{m^2}/\Gamma_m(m)$ .

\subsection{Matrix-variate distributions}

\begin{definition}
\label{defStief}
An $m\times n$ ($m\geq n$) random {\em Stiefel} matrix $\Sm\in \Sc(m,n)$ is such that
$\Sm\Herm\Sm=\Id$ and is uniformly distributed on $\Sc(m,n)$. Then, it has pdf $p(\Sm)=
|\Sc(m,n)|^{-1}$.
\end{definition}
\begin{definition}
\label{defHaar}A square $m\times m$ random unitary matrix
$\Um\in\Uc(m)$ is such that $\Um\Um\Herm=\Um\Herm\Um=\Id$. When it is uniformly distributed on
$\Uc(m)$, it has pdf $p(\Um)=|\Uc(m)|^{-1}$.
\end{definition}

\begin{definition}\label{def:unitarily_invariant}~\cite[Definition 2.6 and Lemma 2.6]{tutorial}
An $m\times m$ Hermitian random matrix $\Am$ is unitarily invariant if the joint distribution of its
entries equals that of $\Vm\Am\Vm\Herm$ where $\Vm$ is any unitary matrix independent of $\Am$.  If
$\Am$ is unitarily invariant, then its eigenvalue decomposition can be written as
$\Am=\Um\Lambdam\Um\Herm$ where $\Um$ is a Haar matrix independent of the diagonal matrix
$\Lambdam$. Since $\Um$ is Haar (isotropic), it is uniformly distributed on $\Uc(m)$.
\end{definition}

\begin{definition}
\label{def:wishart}
Let $\Hm$ be an $m \times n$ matrix whose columns are zero-mean independent complex Gaussian vectors
with covariance matrix $\Thetam$. 
\begin{itemize}
\item For $m\leq n$, the $m \times m$ random matrix $\Wm=\Hm\Hm\Herm$ is a central {complex Wishart}
  matrix, with $n$ degrees of freedom and covariance matrix $\Thetam$ $(\Wm \sim \Wc_m(n,\Thetam))$.
  The joint distribution of the eigenvalues of $\Wm$ coincides with the law of the squared non-zero
  singular values of $\Hm$.  Let $\Wm=\Um\Lambdam\Um\Herm$ be the singular value decomposition (SVD)
  of $\Wm$. If $\Thetam=\Id$, then $\Wm$ is unitarily invariant \cite{tutorial}. In such a case, the joint distribution of the ordered
  eigenvalues $\Lambdam$ can be written as \cite{LozanoNihar,james}
\begin{equation}
\label{eq:pSigma_Rayleigh1}
p(\Lambdam) = \frac{\pi_m^2|\Lambdam|^{n-m}\ee^{-\trace\{\Lambdam\}}}{\Gamma_m(n)\Gamma_m(m)}\Vc^2(\Lambdam)\,.
\end{equation}
\item For  $m>n$,  if the  rows  of $\Hm$  are  independent  and their
  covariance  matrix  is $\Id$,  the
distribution of the ordered eigenvalues of $\Hm\Herm\Hm$ is given by~\cite{LozanoNihar}
\begin{equation}
\label{eq:pSigma_Rayleigh2}
p(\Lambdam) = \frac{\pi_n^2|\Lambdam|^{m-n}\ee^{-\trace\{\Lambdam\}}}{\Gamma_n(m)\Gamma_n(n)} \Vc^2(\Lambdam)\,.
\end{equation}
\end{itemize}
\end{definition}

\begin{definition}
\label{def:non_central_wishart}
 Let $\Hm$ be an $m \times n$ random matrix whose entries are independent, complex, Gaussian random
 variables with unit variance and average $\Mm=\EE[\Hm]$.  Then, matrix $\Wm=\Hm\Hm\Herm$ is
 non-central Wishart~\cite{james}.
\begin{itemize}
\item For $m\leq n$, the distribution of $\Wm$ is given by~\cite[eq. (99)]{james}
\begin{equation}
p(\Wm) = \frac{|\Wm|^{n-m}}{\Gamma_m(n)}\frac{{_0F_1}(\,; n;
  \Mm\Mm\Herm\Wm)}{ \ee^{\trace\{\Wm+\Mm\Mm\Herm\}}} \,.
\label{eq:pW_non_central_wishart}
\end{equation}
If $\Mm\Mm\Herm$ has full rank and distinct eigenvalues, $\mu_1,\ldots,\mu_m$, then the joint pdf
 of the ordered, strictly positive eigenvalues $(\lambda_1,\ldots, \lambda_m)=\diag(\Lambdam)$ of
 $\Wm$ is given by~\cite[eq. (102)]{james}
\begin{eqnarray}
p(\Lambdam) = \frac{|\Lambdam|^{n-m}\Vc(\Lambdam)|\{{_0F_1}(\,; n-m+1; \mu_i\lambda_j)\}|}{(n-m)!^m \ee^{\trace\{\Lambdam+\Mm\Mm\Herm\}}\Vc(\Mm\Mm\Herm)}    \,.
\label{eq:non_central_wishart}
\end{eqnarray}
Note that~\eqref{eq:non_central_wishart} has been obtained from~\cite[eq. (102)]{james} by
exploiting the result in~\eqref{eq:pFq_2arguments}.

As can be observed from~\eqref{eq:pW_non_central_wishart}, if $\Mm\Mm\Herm$ is a scalar matrix
(i.e., $\Mm\Mm\Herm=\mu\Id$), $p(\Wm)$ only depends on the eigenvalues of $\Wm$. Thus $\Wm$ is
unitarily invariant. In such a case, the distribution of $\Lambdam$ can be obtained from 
~\eqref{eq:non_central_wishart} by applying the limit in~\eqref{eq:limit} and the property
in~\eqref{eq:hypergeometric_derivative}, and it is given by
\begin{eqnarray}
p(\Lambdam) =  \frac{\pi_m^2|\Lambdam|^n |\Fm| \Vc(\Lambdam)}{\Gamma_m(m)\Gamma_m(n)\ee^{\mu m+\trace\{\Lambdam\}}} 
\label{eq:non_central_wishart_mu}
\end{eqnarray}
where $(\Fm)_{ij}=\lambda_j^{-i}{_0F_1}(\,; n-i+1; \mu\lambda_j)$.

Note that, since the Vandermonde determinant in (\ref{eq:V_vandermonde}) and pdf are positive by
definition, here and in the following $|\Fm|$ represents the absolute value of the determinant of
matrix $\Fm$. This avoids us to include in the provided results coefficients that account  for the
sign of determinants.
\item For $m>n$, the same expressions as in~\eqref{eq:pW_non_central_wishart},~\eqref{eq:non_central_wishart}, and~\eqref{eq:non_central_wishart_mu} hold but replacing,
  $\Mm$, $m$ and $n$ with, respectively, $\Mm\Herm$, $n$ and $m$.
\end{itemize}
\end{definition}

\begin{lemma}
\label{lemma:2wishart} 
Let $\Hm_1$ and $\Hm_2$ be, respectively, an $m\times n$ and an $m\times p$ ($m\leq p$) Gaussian
complex random matrix whose columns are independent, have zero mean, and covariance $\Thetam_1$
and $\Thetam_2$, respectively.

\begin{itemize}
\item For $m \leq n$, the $m \times m$ random matrix
  $\Wm=(\Hm_2\Hm_2\Herm)^{-1/2}\Hm_1\Hm_1\Herm(\Hm_2\Hm_2\Herm)^{-1/2}$ is a central
  $\Fc$-matrix~\cite{james}. When $\Thetam_1$ and $\Thetam_2$ are both scalar matrices, $\Wm$ is
  unitarily invariant and has a Beta type II distribution~\cite{GuptaNagar}. Specifically, when
  $\Thetam_1\Thetam_2^{-1}=\omega\Id$, the distribution of its ordered eigenvalues is given by
\begin{equation}
p(\Lambdam) =
\frac{\pi^2_m\Gamma_m(p+n)}{\omega^{mn}\Gamma_m(m)\Gamma_m(p)\Gamma_m(n)}\frac{\Vc^2(\Lambdam)|\Lambdam|^{n-m}}{|\Id+\Lambdam/\omega|^{p+n}}
\,.
\label{eq:IL_Rayleigh1}
\end{equation}

\item  For $m>n$, the matrix $\Wm = \Hm_1\Herm(\Hm_2\Hm_2\Herm)^{-1}\Hm_1$ is unitarily invariant and
  the distribution of its ordered eigenvalues can be expressed as
\begin{equation}\label{eq:IL_Rayleigh2}
p(\Lambdam) 
=  \frac{\pi_n^2\Gamma_m(p\plus n)|\Fm||\Omegam|^{-n}\Vc(\Lambdam)|\Id\plus \Lambdam|^{m\minus p \minus n\minus 1}}{(p\plus n\minus m)!^{-n}\Gamma_n(p\plus n)\Gamma_m(p)\Gamma_n(n)\Vc(\Id \minus \Omegam^{-1})}
\end{equation}
where $\Omegam=\Thetam_1^{1/2}\Thetam_2^{-1}\Thetam_1^{1/2}$, $(\Fm)_{ij}={_1F_0}(p+n-m+1;
;(1-\omega_i^{-1})\lambda_j/(1+\lambda_j))$ for $i=1,\ldots,m$, $j=1,\ldots,n$, and
$(\Fm)_{ij}=(1-\omega_i^{-1})^{m-j}$ for $i=1,\ldots,m$,
$j=n+1,\ldots,m$.  
\end{itemize}
\end{lemma}
\begin{IEEEproof}
The proof is given in Appendix~\ref{app:2wishart}.
\end{IEEEproof}

\begin{lemma}
\label{def:non_central_2wishart}
Let $\Hm_1$ and $\Hm_2 $ be, respectively, an $m\times n$ and an $m\times p$ Gaussian complex random
matrix whose columns are independent and have covariance $\Thetam$. Let also $\EE[\Hm_1]=\Mm$ and
$\EE[\Hm_2]=\zerov$.

\begin{itemize}
\item For $m \leq n$, $\Mm\Mm\Herm=\mu\Id$ and $\Thetam=\theta\Id$, the non-central $\Fc$-matrix
$\Wm=(\Hm_2\Hm_2\Herm)^{-1/2}\Hm_1\Hm_1\Herm(\Hm_2\Hm_2\Herm)^{-1/2}$ is unitarily invariant and
the distribution of its eigenvalues is given by
\begin{equation}
p(\Lambdam) 
= \frac{\pi_m^2\Vc(\Lambdam)^2\Gamma_m(p\plus n){_1F_1}(p\plus n;n;\frac{\mu}{\theta}\Lambdam(\Id\plus \Lambdam)^{\minus 1})}{\Gamma_m(m)\Gamma_m(n)\Gamma_m(p)\ee^{\mu m/\theta}|\Lambdam|^{m\minus n}|\Id\plus \Lambdam|^{p\plus n}}
\label{eq:PL_IL_Rice1}
\end{equation}

\item For $m>n$, and $\Mm\Herm\Thetam^{-1}\Mm=\omega\Id$, the matrix
  $\Wm=\Hm_1\Herm(\Hm_2\Hm_2\Herm)^{-1}\Hm_1$ is unitarily invariant and the distribution of its
  eigenvalues is given by
\begin{equation}
p(\Lambdam) = \frac{\pi_n^2\Gamma_n(p\plus n)\ee^{-\omega n}}{\Gamma_n(n)\Gamma_n(m)\Gamma_n(p\plus n\minus m)}
\frac{|\Fm||\Lambdam|^{m-n}\Vc(\Lambdam)}{|\Id\plus \Lambdam|^{p+1}}
\label{eq:PL_IL_Rice2_omega}
\end{equation}
where $(\Fm)_{ij}=(\lambda_j/(1+\lambda_j))^{n-i}{_1F_1}(p+n-i+1; m-i+1; \omega\lambda_j/(1+\lambda_j))$.
\end{itemize}

\end{lemma}
The proof is provided in Appendix~\ref{app:IL_Rice1}.

\begin{definition}
\label{def:beta}
The $n \times n$ random matrix $\Bm$ is Beta-distributed with positive integer parameters $p$ and
$q$ $(\Bm\sim \Bc_n(p,q))$ if 
\begin{itemize}
  \item given $\Tm$ an upper triangular matrix with positive
    diagonal elements, we can write $\Bm=(\Tm\Herm)^{-1}\Cm\Tm$ where $\Cm\sim \Wc_n(p,\Thetam)$, and 
  \item given $\Am \sim \Wc_n(m,\Thetam)$, we can write $\Am+\Cm=\Tm\Herm\Tm$. Notice that, if either $p<n$ or
    $q<n$, or both $p<n$ and $q<n$, the distribution is referred to as \emph{pseudo}-Beta since it involves \emph{pseudo}-Wishart
    matrices~\cite[and references therein]{durisi}.
\end{itemize}
\end{definition}

When $n \leq p$, $\Bm$ admits an eigendecomposition where the matrix of the eigenvectors is independent of the matrix
of the eigenvalues~\cite[Lemma 8]{durisi}.

\begin{itemize}
  \item For $q\leq n$, the distribution of the $q$ ordered non-zero eigenvalues of
    $\Bm$ is given by~\cite[eq. (13)]{durisi}:
\begin{equation}
\label{eq:beta2}
 p(\Lambdam) = \frac{\pi_q^2\Gamma_q(p+q) |\Id-\Lambdam|^{n-q}|\Lambdam|^{p-n}\Vc^2(\Lambdam)}{\Gamma_q(n)\Gamma_q(p+q-n)\Gamma_q(q)}\,. 
\end{equation}
 \item For $q > n$, $\Bm$ has $n$ nonzero eigenvalues, whose ordered joint
    distribution is given by~\cite[eq. (12)]{durisi}:
\begin{equation}
\label{eq:beta1}
p(\Lambdam)=\frac{\pi_n^2\Gamma_n(p+q)|\Id-\Lambdam|^{q-n}|\Lambdam|^{p-n}\Vc^2(\Lambdam)}{\Gamma_n(n)\Gamma_n(p)\Gamma_n(q)}
\,.
\end{equation}
\end{itemize}

Due to the lack of the corresponding expression in the literature, herein we derive the expression
of the marginal distribution of a single unordered eigenvalue of a $\Bc_n(p,q)$-distributed
matrix, which will be needed in our subsequent derivations.

\begin{proposition}\label{prop:marginal}
Given an $n \times n$ matrix $\Bm\sim \Bc_n(p,q)$, 
\begin{itemize}
\item For $q\leq n$, the pdf of a single unordered eigenvalue of $\Bm$ 
  is given by
\begin{eqnarray}
\label{eq:beta_eig2}
  p(\lambda)
&=&\frac{\pi_q^2}{q\Gamma_q(q)}\frac{\Gamma_q(p+q)\Gamma(n-q+1)}{\Gamma_q(n)\Gamma_q(p+q-n)}
  \non
&& \quad \cdot \sum_{i,j=1}^n \lambda^{p-n+i+j-2}(1\minus \lambda)^{n-q}\Dc_{ij}
\end{eqnarray}
with  $\Dc_{ij}$ being the $(i,j)$-cofactor of the ($n \times n$) matrix $\Am$ such that
\begin{equation}
\label{eq:a_single2} 
(\Am)_{\ell k}= \frac{\Gamma(p-n+\ell+k-1)}{\Gamma(p+k-q+\ell)}\,.
\end{equation}
\item For $q> n$, the pdf of a single unordered eigenvalue of $\Bm$ is given by
\begin{eqnarray}
\label{eq:beta_eig1}
p(\lambda)
&=&\frac{\pi_n^2}{n\Gamma_n(n)}\frac{\Gamma_n(p+q)\Gamma(q-n+1)}{\Gamma_n(p)\Gamma_n(q)}
 \non
&& \quad \cdot \sum_{i,j=1}^n \lambda^{(p-n+i+j-2)}(1\minus \lambda)^{q-n}\Dc_{ij}
\end{eqnarray}
with $\Dc_{ij}$ being the $(i,j)$-cofactor of the ($n \times n$) matrix $\Am$ such that
\begin{equation}
\label{eq:a_single1} 
(\Am)_{\ell k}= \frac{\Gamma(p-n+\ell+k-1)}{\Gamma(p+k+q-2n+\ell)}\,.
\end{equation}
\end{itemize}
\end{proposition}

\begin{IEEEproof}
The proof is given in Appendix~\ref{app.A}.
\end{IEEEproof}

\subsection{Communication model}

We consider a  single-user multiple-antenna communication system, with $m$  and $n$ denoting the
number  of receive  and  transmit  antennas, respectively.   Assuming  block-memoryless fading  with
coherence length equal  to $b$, the  output can be described by  the following linear
relationship:
\begin{equation}
\Ym = \sqrt{\gamma}\Hm \Xm+\Nm
\label{lin-model}
\end{equation}
where $\Ym$ is the $m \times b$ output matrix, and $\Hm$ is the $m\times n$ complex random channel
matrix whose entries represent the fading coefficients between each transmit and receive
antenna. $\Nm$ is the $m \times b$ matrix of white Gaussian noise which is assumed to have
i.i.d. complex Gaussian entries with zero mean and unitary variance.  The normalized per-transmit
antenna SNR is denoted by $\gamma={\rm SNR}/n$, and $\Xm$ is the random complex $n \times b$ input
matrix whose structure will be specified in the following sections.
Moreover, for any positive integer $n$, we define
\[ \gamma_n=\gamma^{n(n-1)/2}\,. \]

Note that the above communication model is adopted in all the
following sections, except for Section~\ref{subsec:Interference} where we resort to a slightly
different model explicitly accounting for interference.

\section{Output Statistics with IID Gaussian
  Input\label{sec:analysis1}}

In this section, we analyse the case where the distribution of $\Xm$ is Gaussian i.i.d.  and
consider both the noise-limited and interference-limited scenarios. Note that, in the case under
study, the average energy of the input signal is given by $\EE[\trace\{\Xm\Xm\Herm\}] = nb$.

As for the communication channel, we focus our analysis on some classes of channel matrices whose
Gramian $\Wm=\Hm\Hm\Herm$ is unitarily invariant. As shown in the following, this allows us to write
the expression of the output pdf in terms of the distribution of the eigenvalues of the channel
matrix.  In particular, in both the noise-limited and the interference-limited case, we draw on the
following results:

\begin{itemize}
\item for $m\le n$, and for unitarily invariant $\Hm\Hm\Herm$, the distribution of
  $\Ym$ is given by~\cite[eq. (40) and (41)]{LozanoNihar}
\begin{equation}\label{eq:pY_pSigma1}
p(\Ym)= \frac{\Gamma_{m}(m)\ky }{\pi_m\gamma_m}\int\frac{|\Em||\Id+\gamma\Lambdam|^{m-b-1}}{\Vc(\Lambdam)}p(\Lambdam) \dd\Lambdam\,,
\end{equation}
where $\Lambdam$ is an $m\times m$ diagonal matrix containing the
eigenvalues of channel matrix  $\Hm\Hm\Herm$,
$(\Em)_{ij}=\ee^{y_ic_j}$, and $c_j=\gamma\lambda_j/(1+\gamma\lambda_j)$, $j=1,\ldots, m$. Moreover,
$y_1,\ldots,y_{m}$ are the eigenvalues of $\Ym\Ym\Herm$ and  
\begin{equation}
K(\Ym) = \frac{\ee^{-\|\Ym\|^2}}{\pi^{mb}\Vc(\Ym\Ym\Herm)}\,. 
\label{eq:ky}
\end{equation}
\item for $m > n$, and for unitarily invariant $\Hm\Herm\Hm$, the pdf of $\Ym$ can be
  obtained by following the steps described in~\cite{LozanoNihar} and is given by
\begin{equation}
p(\Ym) =  \frac{\Gamma_{n}(m) \ky }{\pi_n\gamma^{n(m-n)}}
\int \frac{|\widetilde{\Em}||\Id+\gamma\Lambdam|^{m-b-1}}{\Vc(\gamma\Lambdam)|\Lambdam|^{m-n}}
p(\Lambdam) \dd\Lambdam
\label{eq:pY_pSigma2}
\end{equation}
where $\widetilde{\Em}$ is an $m\times m$ matrix whose elements are given by $(\widetilde{\Em})_{ij}
= \ee^{y_i c_j}$ for $1\le j\le n$, and $(\widetilde{\Em})_{ij} = y^{j-n-1}_i$ for $n+1 \le j\le
m$.
 Note that in this case the matrix $\Hm\Hm\Herm$ is of reduced rank since it has $m-n$ zero
eigenvalues. Thus, here $p(\Lambdam)$ indicates the distribution of the $n$ non-zero eigenvalues
of $\Hm\Hm\Herm$ and $\Lambdam$ is an $n\times n$ diagonal matrix. 

\begin{IEEEproof}
The proof is given in Appendix~\ref{app:main_result}.
\end{IEEEproof}
\end{itemize}

\subsection{Noise-limited\label{subsec:NoiseLimited}}
The output pdf of the uncorrelated Rayleigh-faded channel has been evaluated in
\cite{LozanoNihar}. For sake of completeness, we recall this result and present the corrected expression of the output
pdf when $m> n$.  Then, we extend the analysis to two other practically relevant fading models,
namely, the Rician block-fading channel \cite{Giulio:07,TariccoColuccia} and Land Mobile Satellite
(LMS) channel \cite{lmsmimo,Liolis}.

\subsubsection{Rayleigh fading channel}
In the case of uncorrelated Rayleigh channel, the entries of $\Hm$ follow an i.i.d. zero-mean,
unit-variance, complex Gaussian distribution.
\begin{itemize}
\item For $m \leq n$, the distribution of the eigenvalues of $\Hm\Hm\Herm$ is given
  by~\eqref{eq:pSigma_Rayleigh1}.  It follows that, by using~\eqref{eq:pY_pSigma1} and the result in
  Appendix~\ref{app:integral_2determinants}, the distribution
  of $\Ym$ can be written as~\cite[Proposition 2]{LozanoNihar}.
\begin{equation}\label{eq:Rayleigh1}
p(\Ym)=\frac{\pi_m }{\gamma_m\Gamma_{m}(n)}\ky |\Zm|
\end{equation}
where the $i,j$-th entry of the $m\times m$ matrix $\Zm$ is given by
\[ (\Zm)_{ij}= \int_0^{\infty}\exp\left(\frac{y_i\gamma x}{1+\gamma x}-x\right)\frac{x^{n-m+j-1}}{(1+\gamma x)^{b+1-m}}\dd x\,. \]

\item For $m > n$, the distribution of the eigenvalues of channel
  matrix $\Hm\Herm\Hm$ is given
  by~\eqref{eq:pSigma_Rayleigh2}.  By applying~\eqref{eq:pY_pSigma2} and the result in
  Appendix~\ref{app:integral_2determinants}, the output pdf is given by
\begin{eqnarray}\label{eq:Rayleigh2}
p(\Ym)= \frac{\pi_n }{\Gamma_{n}(n)\gamma^{n(m-n)}}\ky |\Zm| \,.
\end{eqnarray}
Note that the expression above differs from the one presented in~\cite[Proposition 2]{LozanoNihar}
in the term $\gamma^{n(m-n)}$, which appears at the denominator.  The
$i,j$-th entry of the $m\times m$ matrix $\Zm$ can be written as
\[ (\Zm)_{ij}=\int_{0}^{\infty}\exp\left(\frac{y_i\gamma x}{1+\gamma x}\minus x\right)
\frac{(x/\gamma)^{j-1}}{(1+\gamma x)^{b+1-m}} \dd x \,, \]
for $1\le i \le m, 1\le j\le n$ and $(\Zm)_{ij}=y_i^{j-n-1}$, for $1\le i \le m, n+1\le j\le m$.
\end{itemize}

\subsubsection{Rician channel}
The Rician channel is traditionally modeled as a superposition of a scattered plus a LOS component, i.e.,
\begin{equation}
\Hm=\sqrt{\frac{\kappa}{\kappa+1}} \bar{\Hm}+
\sqrt{\frac{1}{\kappa+1}} \widetilde{\Hm} \,.
\label{eq:ricean_channel}
\end{equation}
In (\ref{eq:ricean_channel}), $\kappa$ is the Rician factor representing the ratio of the average
power of the unfaded channel component to the faded channel component, the entries of
$\widetilde{\Hm}$ are independent, zero-mean unit-variance complex Gaussian, and $\bar{\Hm}$ is a
deterministic matrix representing the LOS component.

Specifically, for $m \leq n$, we consider the special case $\bar{\Hm}\bar{\Hm}\Herm = h\Id$ (for $m
> n$ we assume $\bar{\Hm}\Herm\bar{\Hm} = h\Id$), where $h$ is a positive parameter.  This
assumption reflects two main settings: the scalar LOS channel, introduced in \cite{Mosscalar} and
therein already analysed in the high-SNR regime, and the LOS MIMO with residual scattering
\cite{Oien}.  Both models assume the LOS matrix to have high (full) rank. The one in \cite{Oien} is
suitable for MIMO backhaul links where antenna spacing is carefully designed and transmit-receive
distance is fixed. Our model can be thought of as a Gaussian perturbation, with small variance, of
the one in \cite{Oien}.  The model in \cite{Mosscalar}, although being a sub-case of the one in
\cite{Oien} from the pure mathematical viewpoint, has played a major role in the early
characterization of MIMO Rician channels, due to the amenability of diagonal \cite{hosli} (and, in
particular, scalar) non-centrality matrices for the derivation of the capacity-achieving input law.

Under the aforementioned assumption, the Gramian of the matrix $\Hm$ is unitarily invariant (see
Definition~\ref{def:non_central_wishart}), thus the pdf of the output can be expressed as in the
following proposition.

\begin{proposition}\label{GaussianRician}
Given a channel as in~\eqref{lin-model} and~\eqref{eq:ricean_channel}, with i.i.d. Gaussian input
and Rician block-fading,
\begin{itemize}
\item for $m \leq n$, and $\bar{\Hm}\bar{\Hm}\Herm = h\Id$, the pdf of its output can be written as
 \begin{equation}
 p(\Ym)=\frac{\pi_m(1+\kappa)^{mn}}{\gamma_m\Gamma_m(n)\ee^{\kappa hm}}
 \ky |\Zm|
\,,
 \label{eq:pYRice1}
\end{equation}
 where \[
 (\Zm)_{ij}=\int_{0}^{\infty}\frac{\ee^{y_i\gamma x/(1+\gamma x)}{_0F_1}(\,;n\minus j\plus 1;\hat{x})\dd x}{\ee^{(1\plus \kappa)x}x^{j-n}(1\plus \gamma x)^{b-m+1}}
 \]

\item for $m > n$, and $\bar{\Hm}\Herm\bar{\Hm} = h\Id$, the following result holds
\begin{equation}
\label{eq:pYRice2}
p(\Ym)= \frac{\pi_n(1+\kappa)^{nm}}{\gamma_n\gamma^{n(m-n)}\Gamma_n(n)\ee^{\kappa hn}} \ky |\Zm|
\end{equation}
where 
\[ (\Zm)_{ij}= \int_0^{\infty} \frac{\ee^{y_i\gamma x/(1+\gamma x)}{_0F_1}(\,; m\minus j \plus 1; \hat{x})\dd x}{\ee^{(1+\kappa)x}x^{j-n}(1+\gamma x)^{b-m+1}}\,, \] 
for $1\le i \le m, 1\le j\le n$ and $(\Zm)_{ij}=y_i^{j-n-1}$, for $1\le i \le m, n+1\le j\le m$,
\end{itemize}
with $\hat{x} = \kappa(1\plus \kappa) hx$.
\end{proposition}

\begin{IEEEproof}
The proof is given in Appendix~\ref{app:fading_rice}.
\end{IEEEproof}

\subsubsection{Land mobile satellite communication}
The Land  Mobile Satellite (LMS) MIMO  channel can be viewed as  a non-central channel
  with random mean. Thus,  the channel matrix model can be described as
\begin{equation}
 \label{eq:LMS_channel}
 \Hm = \bar{\Hm}+\widetilde{\Hm} 
\end{equation} 
  where the entries of $\widetilde{\Hm}$ are independent, zero-mean unit-variance complex Gaussian
  and $\bar{\Hm}$ is a random matrix.  As shown in~\cite{lmsmimo}, in a LMS channel the matrix
  $\bar{\Hm}\bar{\Hm}\Herm$ follows a matrix-variate $\Gamma(\alpha,\Omegam)$
  distribution~\cite{Nagar} where $\alpha$ plays the role of a shape parameter, while $\Omegam$ is a
  scale parameter.  Indeed, $\alpha$ can be viewed as a generalized number of degrees of freedom of
  the non-centrality parameter, while $\Omegam$ is related to the average power of the random LOS
  component, as discussed in detail in \cite{lmsmimo}. Assuming $\Omegam=\omega\Id$, $\Hm\Hm\Herm$
  is unitarily invariant, as shown in \cite[Property 1]{lmsmimo}.

  Under this assumption, the expression of the output pdf can be expressed as in the following
  proposition.

\begin{proposition}\label{GaussianLMS}
Given an LMS MIMO channel as in~\eqref{eq:LMS_channel} with $\Omegam=\omega\Id$, 
\begin{itemize}
\item  for $m\leq n$, the pdf of its output can be written as
\begin{equation}\label{eq:pY_LMS1}
p(\Ym)= \frac{\pi_m}{\gamma_m\Gamma_m(n)(1+1/\omega)^{m\alpha}}\ky |\Zm|\,,
\end{equation}
 where
 \[
 (\Zm)_{ij}=\int_{0}^{\infty}
 \frac{\ee^{y_i\frac{\gamma x}{1+\gamma x}}{_1F_1}\left(\alpha\minus j\plus 1;n\minus j\plus 1;\frac{x}{1\plus \omega}\right)}{\ee^x x^{j-n}(1\plus \gamma x)^{b-m+1}} 
 \dd x\,;
 \]

\item for $m > n$, the output pdf is given by:
\begin{equation}\label{eq:pY_LMS2}
p(\Ym) = \frac{\pi_n}{\gamma_n\gamma^{n(m-n)}\Gamma_n(m)  (1+1/\omega)^{n\alpha}}\ky |\Zm|
\end{equation}
 where
 \[ (\Zm)_{ij}=\int_{0}^{\infty}\frac{\ee^{y_i\frac{\gamma x}{1+\gamma x}}{_1F_1}\left(\alpha\minus j\plus 1;m\minus j\plus 1;\frac{x}{1\plus \omega}\right)}{\ee^x x^{j-n}(1+\gamma x)^{b-m+1}}  
 \dd x\,,\]
for $1\le i \le m, 1\le j\le n$ and $(\Zm)_{ij}=y_i^{j-n-1}$, for $1\le i \le m, n+1\le j\le m$. 
\end{itemize}

\end{proposition}

\begin{IEEEproof}
The proof is given in Appendix~\ref{app:LMS}.
\end{IEEEproof}

\subsection{Interference limited\label{subsec:Interference}}
We now consider the case where  the  main  impairment   to  communication  is  represented   by  the  co-channel
  interference. In particular, each  interferer is seen from the direct link  receiver under its own
  random channel, which we assume to be  affected by Rayleigh fading, again with block-length $b$.
  We assume that  there are $L$ active  interferers in the network, each  equipped, for homogeneity,
  with the same  number of antennas, $n$, as  the transmitter of the
  useful signal.  We evaluate  the output pdf
  when a whitening filter  is applied to the received signal and we  consider two channel models. In
  the  former, the  desired signal  undergoes Rayleigh  fading;  in the  latter, the  direct link  is
  affected  by Rician  fading, i.e.,  we assume  the existence  of an  LOS path  between  the useful
  transmitter and its intended receiver.

The received signal can be modeled as
\begin{eqnarray}
\widetilde{\Ym}=\sqrt{\gamma}\Hm_s\Xm+\Wm
\label{eq:interference_limited}
\end{eqnarray} 
where 
\[ \Wm=\sum_{\ell=1}^{L}\widehat{\Hm}_{\ell}\widehat{\Xm}_{\ell}\]
represents the interference. Specifically, the $m\times n$ matrix $\widehat{\Hm}_{\ell}$ models
the channel connecting  the $\ell$-th interferer with the receiver, while  the $n\times b$ matrix
$\widehat{\Xm}_{\ell}$   represents   the   signal   transmitted  by   the   $\ell$-th   interferer,
$\ell=1,\ldots,  L$.  The interference  can be  rewritten as  $\Wm=\widehat{\Hm}\widehat{\Xm}$ where
$\widehat{\Hm}=[  \widehat{\Hm}_1,\ldots,  \widehat{\Hm}_L]$  is  an  $m\times  Ln$  matrix  and
$\widehat{\Xm}=[\widehat{\Xm}_1\Herm, \ldots,
\widehat{\Xm}_L\Herm]\Herm$ is of size $Ln\times b$.
By assuming that the entries of $\widehat{\Xm}$  are i.i.d. complex Gaussian with zero mean and unit
variance, the covariance of the interference,  conditioned on the knowledge of the composite channel
matrix $\widehat{\Hm}$, is given by
\[  \Rm=\EE[\Wm\Wm\Herm | \widehat{\Hm}]=
\widehat{\Hm}\EE[\widehat{\Xm}\widehat{\Xm}\Herm]\widehat{\Hm}\Herm =
b\widehat{\Hm}\widehat{\Hm}\Herm \,.\] 

We apply to the received signal $\widetilde{\Ym}$ the whitening filter $\Bm=\sqrt{b}\Rm^{-1/2}$ and obtain
\begin{eqnarray}
\Ym 
&=& \Bm\widetilde{\Ym} \non
&=& \sqrt{b}\Rm^{-1/2}\widetilde{\Ym} \non
&=& \left(\widehat{\Hm}\widehat{\Hm}\Herm\right)^{-1/2} (\sqrt{\gamma}\Hm_s\Xm+\Wm) \non
&=& \sqrt{\gamma}\Hm\Xm+ \Nm \label{eq:whitened}
\end{eqnarray}
where $\Hm=\left(\widehat{\Hm}\widehat{\Hm}\Herm\right)^{-1/2}\Hm_s$ and $\Nm=\left(\widehat{\Hm}\widehat{\Hm}\Herm\right)^{-1/2}\Wm$. Clearly, $\EE[\Nm\Nm\Herm | \widehat{\Hm}] = b\Id$.
In the following, we provide the pdf of $\Ym$.

\subsubsection{Rayleigh fading channel}
\begin{proposition}\label{GaussianIL}
We consider the interference-limited channel described by~\eqref{eq:interference_limited}, with $L$
active interferers, i.i.d. Gaussian input and Rayleigh fading. 
If $\Hm_s\Hm_s\Herm \sim
\Wc_{m}(n,\Thetam_s)$ and $\widehat{\Hm}\widehat{\Hm}\Herm \sim
\Wc_{m}(Ln,\widehat{\Thetam})$, then we have the following results. 

\begin{itemize}
\item For $m \leq n$, due to mathematical constraints, we only analyse the case of spatially
  uncorrelated receiving antennas, i.e., $\Thetam_s=\theta_s\Id$ and
  $\widehat{\Thetam}=\hat{\theta}\Id$. Then, the pdf of $\Ym$ can be written as
\begin{equation}\label{eq:PY_IL_Rayleigh1}
p(\Ym)= \frac{\pi_m\Gamma_{m}(Ln+n)}{\gamma_m\omega^{mn}\Gamma_{m}(Ln)\Gamma_{m}(n)}\ky |\Zm|
 \end{equation}
 where $\omega=\theta_s/\hat{\theta}$ and
 \begin{equation}
 (\Zm)_{ij}=\int_{0}^{\infty}
 \frac{\ee^{{y_i\frac{\gamma x}{1+\gamma x}}} x^{n-j}}{(1+\gamma x)^{b-m+1}(1+x/\omega)^{Ln+n}}
 \dd x \,. \nonumber 
 \end{equation}
This result is obtained by substituting~\eqref{eq:IL_Rayleigh1} in~\eqref{eq:pY_pSigma1} and by
exploiting the result in Appendix~\ref{app:integral_2determinants}.

\item For $m > n$, the pdf of $\Ym$ is given by:
\begin{eqnarray}\label{eq:PY_IL_Rayleigh2}
p(\Ym)&=& \frac{\pi_n(Ln\plus n\minus m)!^m \Gamma_{n}(m)\Gamma_m(Ln\plus n)\ky }{\Gamma_n(Ln\plus n)\Gamma_{n}(n)\Gamma_{m}(Ln)\gamma_n \gamma^{n(m-n)}} \non
&&\quad\cdot\frac{|\Omegam|^{m\minus n\minus 1}}{\Vc(\Omegam)}\int
\frac{|\widetilde{\Em}||\Id\plus \gamma\Lambdam|^{m\minus b\minus 1}|\Fm|\dd\Lambdam}{|\Lambdam|^{m\minus n}|\Id\plus \Lambdam|^{Ln\plus n\minus m\plus 1}}\non
 \end{eqnarray}
where $\Omegam=\Thetam_1^{1/2}\Thetam_2^{-1}\Thetam_1^{1/2}$ and the matrices
$\widetilde{\Em}$ and $\Fm$ have been defined below~\eqref{eq:pY_pSigma2}
and~\eqref{eq:IL_Rayleigh2}, respectively. This result is obtained by
substituting~\eqref{eq:IL_Rayleigh2} in~\eqref{eq:pY_pSigma2}. However, we cannot solve the integral
by applying the result in Appendix~\ref{app:integral_2determinants} directly. Indeed, although
matrices $\widetilde{\Em}$ and $\Fm$ are both of size $m\times m$, a portion of their columns and
rows is composed of constant terms. Thus, we need to resort to the property of the determinant of
block matrices, in order to obtain $n\times n$ blocks to which the result in
Appendix~\ref{app:integral_2determinants} can be applied. We skip the details of this procedure due
to the cumbersome expressions that are involved.
\end{itemize}
\end{proposition}

\subsubsection{Rician fading channel}

\begin{proposition}\label{RiceIL}
We consider the interference-limited channel described by~\eqref{eq:interference_limited}, with $L$
active interferers, i.i.d. Gaussian input, Rician faded useful signal and Rayleigh fading affecting
the interfering links.  For a Rician channel, matrix $\Hm_s$ can be written as
in~\eqref{eq:ricean_channel}
\[ \Hm_s = \sqrt{\frac{\kappa}{\kappa+1}} \bar{\Hm}_s+ \sqrt{\frac{1}{\kappa+1}} \widetilde{\Hm}_s  \]
where $\kappa$ is the Rician factor, $\bar{\Hm}_s$ is deterministic, and $\widetilde{\Hm}_s$ is
complex Gaussian with independent colums whose covariance is
$\Thetam$.
According to our assumptions on LOS links made in
Section~\ref{subsec:NoiseLimited}, we have: 
\begin{itemize}
\item for $m\leq n$,  setting  $\Thetam_s=\widehat{\Thetam}=\theta\Id$
  and $\bar{\Hm}_s\bar{\Hm}_s\Herm=h\Id$, 
\begin{equation}
\label{eq:PY_IL_Rice1}
p(\Ym)= \frac{\pi_m\Gamma_m(Ln+n)\ee^{-h\kappa m/\theta} }{\gamma_m\Gamma_m(n)\Gamma_m(Ln)\tilde{\kappa}^{-mn}}\ky |\Zm|
 \end{equation}
 where 
\begin{equation}
 (\Zm)_{ij}=\int\displaylimits_{0}^{\infty}
 \frac{\ee^{y_i\frac{\gamma x}{1+\gamma x}}
  {_1F_1}(\tilde{L}\plus j;n\minus m\plus j;h\kappa\tilde{\kappa}\tilde{x}/\theta) \dd x}{(1\plus\gamma
   x)^{b-m+1}(1\plus \tilde{\kappa}x)^{\tilde{L}+1} x^{m-n}\tilde{x}^{1-j}}
 \end{equation}
with $\tilde{\kappa} = 1+\kappa$, $\tilde{L}=Ln+n-m$, and $\tilde{x} = x/(1+\tilde{\kappa}x)$
\item for $m > n$, and $\bar{\Hm}_s\Herm\Thetam^{-1}\bar{\Hm}_s=h\Id$,  
\begin{equation}\label{eq:PY_IL_Rice2}
p(\Ym)= \frac{\pi_n\Gamma_n(Ln\plus n)\tilde{\kappa}^{nm}\ee^{-h\kappa n} }{\gamma_n\gamma^{n(m-n)}\Gamma_n(n)\Gamma_n(Ln\plus n\minus m)}
\ky |\Zm|
 \end{equation}
where
\[ (\Zm)_{ij}=\int\displaylimits_0^{\infty}  \frac{ {_1F_1}(Ln\plus n\minus j\plus 1; m\minus j\plus 1; h\kappa\tilde{\kappa}\tilde{x})\dd x}{\ee^{\frac{-y_i\gamma x}{1+\gamma x}}(1+\tilde{\kappa}x)^{Ln+1}\tilde{x}^{j-n}(1\plus \gamma x)^{b-m+1}}, \]
for $1\le i \le m, 1 \le j\le n$, and $(\Zm)_{ij}=y_i^{j-n-1}$, for $1\le i \le m, n+1\le j\le m$\,;
with $\tilde{\kappa} = 1+\kappa$ and $\tilde{x} =
x/(1+\tilde{\kappa}x)$.
\end{itemize}

\end{proposition}

\begin{IEEEproof}
The proof is given in Appendix~\ref{app:PY_IL_Rice}.
\end{IEEEproof}

Note that also in this case mathematical issues made the analysis only possible for uncorrelated
receivers.

\subsection{{Exploitation of the analytical results}}
The mutual information between the channel input, $\Xm$, and the channel output, $\Ym$,
normalized to the fading coherence length, can be expressed as:
\begin{equation}\label{MI}
 \Ic =\frac{1}{b}\left[ h(\Ym)- h(\Ym|\Xm)\right]\,
\end{equation}
where $h(\Ym)=\EE[-\log p(\Ym)]$ and $h(\Ym|\Xm)=\EE[-\log p(\Ym|\Xm)]$.  Once the pdf of the
channel output, $p(\Ym)$, is obtained, it can be used to evaluate its differential entropy,
$h(\Ym)$.  For Rayleigh and Gaussian channels with identity covariance matrix, considering that
$\Xm$ is given, the output $\Ym$ is complex Gaussian and its rows are i.i.d. Hence, in order to
derive the conditional differential entropy $h(\Ym|\Xm)$, we can compute its value for an arbitrary
row of $\Ym$ and then scale it by the number of rows of $\Ym$ \cite{LozanoNihar}.

In~\cite{LozanoNihar}, the mutual information has been
computed in presence of Rayleigh channel and i.i.d. Gaussian input,
for $m\le n$. In the following, we provide three examples
of mutual information computation. First, we address the case of
noise-limited Rayleigh channel with $m>n$ and, then,  the 
noise-limited Rician channel, both with $m\le n$ and  $m>n$.

\begin{figure}
 \centering
  \includegraphics[width=0.95\columnwidth]{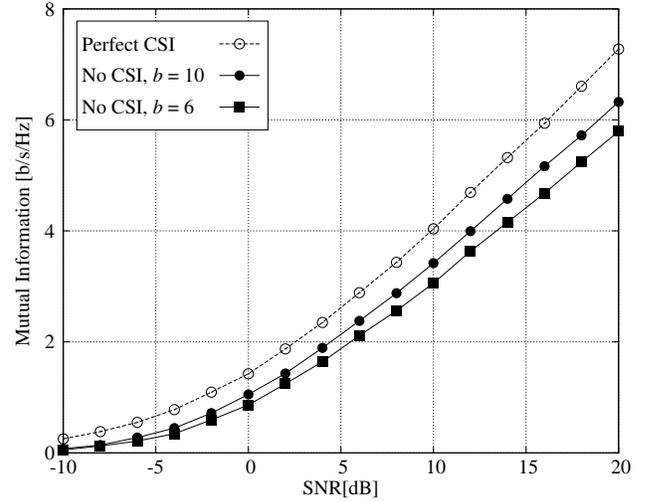}
  \caption{Mutual information vs. SNR in Rayleigh channel: comparison between the case where no CSI
    is available (solid line) and the case of perfect CSI at the receiver (dashed line), with $b=6,
    10$, $m=2$ and $n=1$.}
  \label{Rayleigh}
\end{figure}

In the case of Rayleigh channel, the conditional differential entropy is obtained using
\cite[eq. (4)]{LozanoNihar}, while the unconditional differential entropy is evaluated using
(\ref{eq:Rayleigh1}) or (\ref{eq:Rayleigh2}) depending on the relationship between $m$ and $n$.
Fig.~\ref{Rayleigh} shows the mutual information as a function of the $\mathrm{SNR}$, with $b=6,10$,
$m=2$ and $n=1$, when no channel state information (CSI) is available and in the case of perfect CSI
at the receiver. The latter is obtained by computing \cite[eq. (10)]{LozanoNihar}.  The results
confirm the intuition, as well as previous analysis~\cite{closedform,tse}: the higher the SNR and
the value of $b$, the better the performance, while the lack of CSI causes a noticeable degradation.

For the Rician channel, the expression of the channel matrix is given
by~\eqref{eq:ricean_channel}. By adopting again the method in \cite{LozanoNihar},
the differential entropy of the output conditioned on the input signal can be computed.
Let us denote by $\yv$  an arbitrary row of $\Ym$;   
then, using~\cite[eq. (31)]{LozanoNihar} and considering the
translation-invariant property of differential entropy, we can write
the mutual information when the receiver does not have any knowledge
of the non-LOS component:
\begin{equation}\label{entropyvec}
    h(\yv|\Xm)=h(\yv\Herm|\Xm)=\EE\left[\log_2 \left((\pi \ee)^b\left|\Id\plus \frac{\gamma\Xm\Herm\Xm}{1\plus \kappa} \right|\right)\right]
\end{equation}
with the expectation being over the distribution of $\Xm$. 
The above expression can be conveniently computed resorting to~\cite[eq. (4)]{LozanoNihar}. 
The unconditional differential entropy of the output is derived through (\ref{eq:pYRice1}) and
(\ref{eq:pYRice2}).

\begin{figure}
  \centering
  \includegraphics[width=0.95\columnwidth]{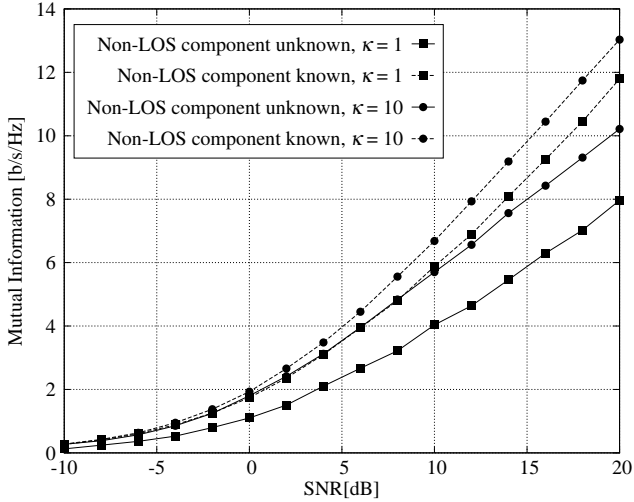}
  \caption{Mutual information vs. SNR in Rician channel: comparison between the case where the
    receiver does not have any knowledge on the non-LOS component (solid line) and when such
    knowledge is available (dashed line), for $b=6$, $n=2$, $m=2$ and $\kappa=1, 10$.}
  \label{Rician}
\end{figure}

Fig.~\ref{Rician} shows the mutual information as a function of the SNR, with $b=6$, $m=2$ and
$n=2$. Rician factors are set to $\kappa=1$ and $\kappa=10$. The plot depicts the mutual information
in the two cases where the receiver has knowledge of the non-LOS component
\cite[eq. (10)]{LozanoNihar} and where it does not \eqref{entropyvec}. The deterministic channel
matrix in (\ref{eq:ricean_channel}) is set as follows:
\[
\bar{\Hm}= \left[ \begin{array}{cc} \sqrt{2} & 0\\ 0 & \sqrt{2}
\end{array} \right] .
\]
In Fig.~\ref{Rician}, the relative gap between the achievable mutual information in the two
scenarios with $\kappa=1$ is more evident than for $\kappa=10$, since the higher the Rician factor,
the higher the amount of information on the LOS component, which is known at the receiver. This is
also compliant with the monotonicity results in~\cite{hosli}.

\begin{figure}
  \centering
  \includegraphics[width=0.95\columnwidth]{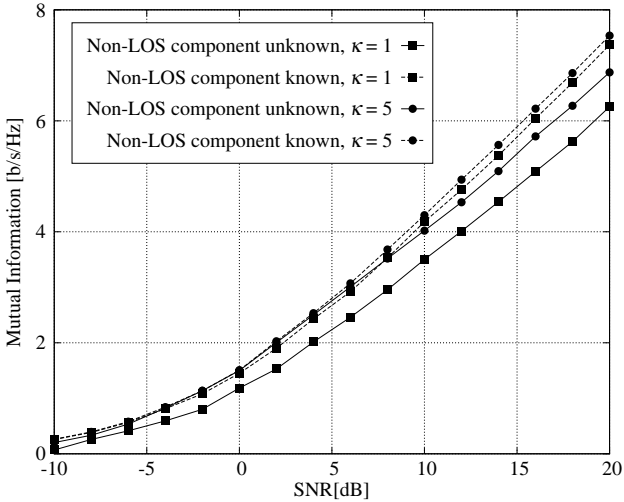}
  \caption{Mutual information vs. SNR in Rician channel: comparison between the cases where
    knowledge of the non-LOS component is not available at the receiver (solid line) and when it is
    (dashed line). $b=6$, $m=2$, $n=1$ and $\kappa=1, 5$.}
  \label{Rician2}
\end{figure}

Finally, Fig.~\ref{Rician2} shows the mutual information for the two scenarios above, in the case of
$m>n$, namely, $m=2$, $n=1$, and $b=6$.  The Rician factor is set to $\kappa=1$ and $\kappa=5$.  In
this scenario, the deterministic channel matrix is set to $\bar{\Hm}=\left[ \sqrt{3/2}, 1/\sqrt{2}
  \right]^H$.  Similar observations to those above hold. However, comparing Fig.~\ref{Rician} to
Fig.~\ref{Rician2}, we notice that, as expected, the reduction in the number of antennas at the
transmitter leads to severe performance degradation.

\section{Output Statistical Characterization with Product Input Form \label{sec:analysis2}}

As in \cite{LozanoNihar,tse,hochwald,lapidoth:03}, we assume total lack of CSI at both the ends of
the wireless link.  This case is of particular interest for the energy efficiency of the
communication, as the availability of CSI would imply a high energy and time consumption at both the
transmitter and the receiver.  Under this assumption, in the high-SNR regime, the capacity-achieving
input matrix $\Xm$ is proven to have a product structure~\cite[theorem 2]{hochwald} and can be written as
\begin{equation} 
\Xm=\sqrt{c}\Dm^{1/2}\Phim
\label{eq:XD}
\end{equation}
where $c$ is a normalizing constant and $\Dm$ is a real random $n\times n$ diagonal matrix, which is
positive definite with probability $1$. The entries of $\Dm$ represent the amount of transmit power
allocated to each of the $n$ transmit antennas, while $\Phim \in \Sc(n,b)$ represents the
beamforming $n\times b$ matrix. In order to be consistent with the definition of SNR,
we impose the constraint on the average input energy $\EE[\trace\{\Xm\Xm\Herm\}]=nb$.  It follows
that, for our specific input structure, the normalizing constant is given by:
\begin{equation}
c = \frac{nb}{\EE[\trace\{\Dm\}]} \,.
\label{eq:constant_product_form}
\end{equation}

In~\cite[Lemma 10]{durisi}, it is proven that without CSI, in Rayleigh block-fading channels, the
optimal power allocation at the transmitter depends on the relationship between the coherence
length, $b$, and the total number of antennas at both the transmitter and receiver.  Specifically,
\begin{itemize}
\item If $b \ge m+n$, all diagonal entries of $\Dm$ are almost surely
  equal to $1$. This case 
  corresponds to the conventional unitary space-time modulation
  (USTM)~\cite{hochwald}, where  $\Dm=\Id$ and $c=b$;
\item If $b<m+n$, the optimal input is $\Dm\sim \Bc_n(b-n,m+n-b)$, which is referred to as
  Beta-variate space-time modulation (BSTM)~\cite{durisi}.  This scenario allows  the analysis of
  an uplink massive-MIMO system, with $m \geq n$ and even $m\gg n$, which is relevant in the next-generation
  cellular setting. 
\end{itemize}

\subsection{Case $b \geq m+n$}
As mentioned above, when $b\geq m+n$, the optimal power allocation over the transmitter antennas is
given by a diagonal matrix, $\Dm$, with entries almost surely equal to $1$. Under these assumptions,
the following results hold.
\begin{proposition}
\label{uncond} 
Consider a channel as in~\eqref{lin-model}, affected by i.i.d. block-Rayleigh fading 
  and with input given by~\eqref{eq:XD}. Let $\Deltam = \gamma c \Dm(\Id+\gamma c \Dm)^{-1} =
  \diag(\delta_1,\ldots,\delta_n)$, with $\delta_i$'s being distinct values. Then,
\begin{itemize}
\item for $m \leq n$, the pdf of its matrix-variate output, conditioned on $\Dm$ and for $n\leq b$,
  can be expressed as
\begin{equation}
\label{eq:PY|D1}
p(\Ym | \Dm) = \frac{\Gamma_m(b)\ky |\Ym\Ym\Herm|^{m-n}|\Gm|}{\pi_m(b\minus n)!^m   \Vc(\Deltam)|\Id+\gamma c\Dm|^{m}}
\end{equation}
where for $j=1,\ldots,n$
\[ 
(\Gm)_{ij} = \left\{
\begin{array}{ll}
{_1F_1}(1; b\minus n\plus 1; y_i\delta_j) & i=1,\ldots,m \\
\delta_j^{n-i}    & i=m+1,\ldots,n
\end{array}
\right.
\]
\item for $m > n$, the conditioned output pdf becomes 
\begin{equation}
\label{eq:PY|D2}
p(\Ym | \Dm) = \frac{\Gamma_n(b) \ky |\Deltam|^{n-m}|\Gm|}{\pi_n(b\minus m)!^n\Vc(\Deltam)|\Id \plus\gamma c \Dm|^m}
\end{equation}
where $i\mathord{=}1,\ldots,m$, are given by
\[ 
(\Gm)_{ij} = \left\{
\begin{array}{ll}
{_1F_1}(1; b\minus m\plus 1; y_i\delta_j) & j=1,\ldots,n \\
y_i^{m-j}    & j=n+1,\ldots,m \,. 
\end{array}
\right.
\]
\end{itemize}
\end{proposition}

\begin{IEEEproof}
The proof is given in Appendix~\ref{app:uncond}.
\end{IEEEproof}

\subsubsection{Case $\Dm=\Id$}
The expressions of $p(\Ym|\Dm)$ in~\eqref{eq:PY|D1}
and~\eqref{eq:PY|D2} hold provided that the diagonal
elements of $\Dm$ are distinct.
Thus, in general, the unconditional pdf of $\Ym$ can be derived by integrating $p(\Ym|\Dm)$ over the
distribution of $\Dm$.  In this section, however, we focus on a particular power allocation matrix,
$\Dm=\Id$, and, by~\eqref{eq:constant_product_form}, we consider $c=b$.  Note that, in this case the
elements of $\Dm$ are not distinct, and expressions~\eqref{eq:PY|D1} and~\eqref{eq:PY|D2} cannot be
directly evaluated. Indeed, $|\Gm|=0$ and $\Vc(\Deltam)=0$, and again a limit procedure must be
applied.

We first observe that, for $\Dm=\Id$ and $c=b$, we have $\Deltam=\gamma b \Dm (\Id+\gamma b
\Dm)^{-1} = \bar{\delta}\Id$ where $\bar{\delta}= \frac{\gamma b}{1+\gamma b}$.
\begin{itemize} 
\item For $m \leq n$, we apply the limit in~\eqref{eq:limit_n0} to the
  ratio $|\Gm|/\Vc(\Deltam)$ in~\eqref{eq:PY|D1} and, after some algebra, obtain 
\[\lim_{\Deltam \to \bar{\delta}\Id} \frac{|\Gm|}{\Vc(\Deltam)} = 
\frac{\pi_n\Gamma_m(n)(b-n)!^m}{\Gamma_n(n)\Gamma_m(b)} |\Ym\Ym\Herm|^{n-m}|\widehat{\Gm}|\]
where $\widehat{\Gm}$ is an $m\times m$ matrix whose elements are given by
$(\widehat{\Gm})_{ij} = y_i^{m-j}{_1F_1}(n-j+1; b-j+1; y_i\bar{\delta})$, $i=1,\ldots,m$, $j=1,\ldots,m$.
By recalling~\eqref{eq:PY|D1}, the distribution of $\Ym$ is then given by
\begin{eqnarray}
p(\Ym) 
&=& \frac{\pi_n\Gamma_m(n)}{\pi_m\Gamma_n(n)} \frac{\ky |\widehat{\Gm}|}{(1+\gamma b)^{nm}}\,.
\end{eqnarray}
\item For $m > n$, we apply the limit in~\eqref{eq:limit_n0} to~\eqref{eq:PY|D2} and
  obtain
\[ \lim_{\Deltam \to \bar{\delta}\Id} \frac{|\Gm|}{\Vc(\Deltam)} =  \frac{\pi_n(b-m)!^n|\widehat{\Gm}|}{\Gamma_n(b-m+n)} \]
where in this case
\[ (\widehat{\Gm})_{ij} = y_i^{n-j}{_1F_1}(n\minus j\plus 1; b\minus m\plus n\minus j\plus 1; y_i\bar{\delta}) \]
for $i=1,\ldots,m, j=1,\ldots,n$, and $(\widehat{\Gm})_{ij} =y_i^{m-j}$ for $i=1,\ldots,m, j=n+1,\ldots,m$.

By recalling~\eqref{eq:PY|D2}, it follows that
\begin{equation}
p(\Ym) 
= \frac{\Gamma_n(b) \ky \bar{\delta}^{n(n-m)}|\widehat{\Gm}|}{\Gamma_n(b-m+n)(1 \plus\gamma b)^{nm}}\,.
\end{equation}
\end{itemize}

We remark that, under the above assumptions, the output pdf also appears in \cite{closedform}. The
corresponding derivations provided therein involve Fourier integrals and Hankel matrices thus
resulting in a slightly less compact form than ours.

\subsection{A \emph{massive MIMO} regime: $b < m+n$}
Now, we consider  the  case of $b <  m+n$; an instance  of this
scenario, by  letting $m \gg n$,  can adequately model the reverse
link of the 
celebrated massive-MIMO  channel \cite{marzetta-mimo}. 
In presence of uncorrelated  block-Rayleigh fading, the
high-SNR  capacity-achieving input structure,  as already  mentioned, departs  from the  equal power
allocation and is Beta distributed. We provide herein the output pdf for a block-fading channel fed by BSTM \cite{durisi}.

\begin{proposition}\label{unconditional2}
Given a channel as in (\ref{lin-model}), with $\Xm=\sqrt{c}\Dm^{1/2}\Phim$, $n\leq b$, and $\Dm \sim
{\mathcal B}_{n}(b-n,n+m-b)$, the pdf of its output can be written as
\begin{equation}\label{mainout}
p(\Ym)=\frac{\pi_n\Gamma_n(b)\Gamma_n(m)(\gamma c)^{n(n-b)}\ky |\Fm_4||\Zm|}{\gamma_nc^{n(n-1)/2}\Gamma_n(n)\Gamma_n(b\minus n)\Gamma_n(n\plus m\minus b)}
\end{equation}
where $\Zm$ is an $n\times n$ matrix, whose generic entry is given by:
\begin{eqnarray}
(\Zm)_{ij} &=&\int_0^1 \frac{(1-x)^{m-b}x^{i-1-n}}{(1+c\gamma x)^{m-b+1}} \non
&&\quad \cdot \left[ \ee^{y_j\frac{c\gamma x }{1\plus c\gamma  x}}-
\sum_{\ell,k=1}^{b-n}\frac{(\Fm_4^{-1})_{\ell k}}{y_j^{n+k-b}}\ee^{y_{\ell+n}\frac{c\gamma x
    }{1+c\gamma x}} \right]\dd x \non
\end{eqnarray}
with $(\Fm_4)_{ij}=y_{n+i}^{b-n-j}\,i,j=1,\dots, b-n$.
\end{proposition}

\begin{IEEEproof}
The proof is given in Appendix~\ref{app:unconditional2}.
\end{IEEEproof}

\subsection{Exploitation of the analytical results}
We now use the above results to compute the achievable mutual information in a massive MIMO case. In
order to derive the output differential entropy conditioned on the input signal, $h(\Ym|\Xm)$, we
exploit the analytic expression of the conditional pdf of the output, $p(\Ym|\Xm)$, obtained above.

\begin{proposition}\label{cond}
Given a channel as in~\eqref{lin-model}, the differential entropy of the output, $\Ym$, conditioned on  the channel input, $\Xm$, can be written as:
\begin{eqnarray}
h(\Ym|\Xm)&=&bm\log_2(\pi \ee) +Km\sum_{i,j=1}^{n}a_{ij}
\sum_{\ell=0}^{m-b}\frac{(-1)^\ell{m-b \choose \ell}}{s_{i,j,\ell} \minus 1}\non
&&\hspace{-4ex}\cdot\left[\log_2(1\plus c\gamma)\minus\frac{c\gamma{_2F_1}(1, s_{i,j,\ell}; s_{i,j,\ell} \plus 1;\minus \gamma)}{ s_{i,j,\ell}\ln 2} \right] \label{conditional}
\end{eqnarray}
where $K$ is a constant term, $s_{i,j,\ell}=b\minus 2n\plus i\plus j\plus \ell$, and $a_{ij}$ is the $(i,j)$-cofactor of an $n \times n$ matrix $\Am$ such that
\[ \Am_{\ell k}=\frac{\Gamma(b-2n+\ell+k-1) \Gamma(m-n+1)}{\Gamma(b-3n+m+\ell+k)}\,. \]
\end{proposition}

\begin{IEEEproof}
The proof is given in Appendix~\ref{app:cond}. 
\end{IEEEproof}

\begin{figure}
 \centering
  \includegraphics[width=0.95\columnwidth]{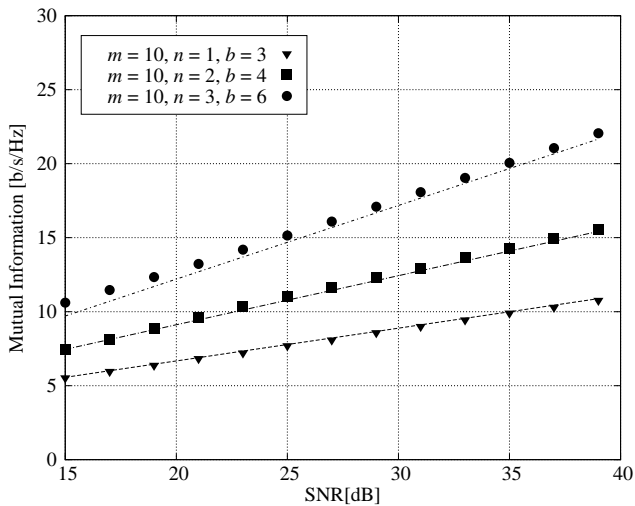}
  \caption{Mutual information vs. SNR in massive MIMO channel with BSTM: $m=10$ and different values
    of $b$ and $n$. Our results (denoted by markers) are compared to the approximation in
    \cite{durisi} (dashed lines).}
  \label{corrR}
\end{figure}

The mutual information obtained in a massive-MIMO-like case is shown in the following figures.
Fig.~\ref{eigR} depicts the mutual information for $n=1$, as the SNR varies and $m$ grows up to very
large values. The plot also compares our results (denoted by markers) are compared to the
approximation given in \cite{durisi} for the high SNR regime (dashed lines). The two sets of curves
match very closely for any value of the parameters, as expected due to the tightness of
\cite[eq. (8)]{durisi}.  As $m$ varies, all three curves have the same slope, as this has been
proven to be insensitive to the number of receiving antennas in our setting
\cite[eq. (8)]{durisi}. As expected, better performance is obtained as $m$ increases. However,
interestingly, Fig.~\ref{corrR} shows that a much higher improvement can be achieved as the fading
coherence length and the number of antennas at the transmitter sightly increase while $m$ is fixed
to 10. In particular, by comparing the two plots, a limited gain in performance is obtained when $m$
increases, while, as expected, the mutual information growth is significant when $n$ is increased by
1.

\begin{figure}
 \centering
  \includegraphics[width=0.95\columnwidth]{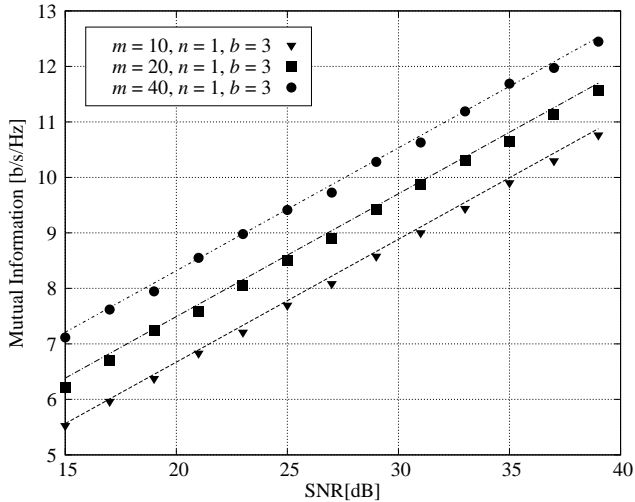}
  \caption{Mutual information vs. SNR in massive MIMO channel with BSTM: $b=3$, $n=1$ and different
    values of $m$. Our results (denoted by markers) are compared to the approximation in
    \cite{durisi} (dashed lines).}
  \label{eigR}
\end{figure}

\section{Conclusion}\label{sec:conclusion}

We obtained new, closed-form expressions for the probability density function of the output signal
of a block-fading MIMO channel. By relying on recent results from the field of finite-dimensional
random matrix theory, we provided results for the case of an i.i.d. Gaussian input under the
assumption that the Gramian of the channel matrix is unitarily invariant. We addressed both the
cases of Rayleigh and Rician fading.  Furthermore, we derived the output probability density
function in the case of product-form input.  We particularized our newly derived expressions to
those already available in the literature for the canonical case of uncorrelated Rayleigh fading,
and we characterized the output signal behavior under different assumptions on the amplitude fading
distribution.

\section{Acknowledgments}
This paper was made possible by NPRP grant $\sharp$5-782-2-322 from the Qatar National Research Fund (a
member of Qatar Foundation). The statements made herein are solely the responsibility of the
authors.

\appendices

\section{Proof of Lemma~\ref{lemma:2wishart}\label{app:2wishart}}
Let $\Hm_1$ and $\Hm_2$ be, respectively, an $m\times n$ and an $m\times p$ ($m\leq
  p$) Gaussian complex random matrix whose columns are independent, have zero mean, and covariance
  $\Thetam_1$ and $\Thetam_2$, respectively. 
\begin{itemize}
\item For $m\leq n$, the distribution of the ordered eigenvalues of
  $(\Hm_2\Hm_2\Herm)^{-1/2}\Hm_1\Hm_1\Herm(\Hm_2\Hm_2\Herm)^{-1/2}$ is given by
  \cite[eq. (98)]{james}
\begin{equation}
p(\Lambdam) = \frac{\pi_m^2\Vc(\Lambdam)|\Lambdam|^{n-m}|\{{_1F_0}(p\plus  n\minus m\plus 1;\,; \minus \frac{\lambda_j}{\omega_i})\}|}{(p\mathord{+}n\mathord{-}m)!^{-m}\Gamma_m(p)\Gamma_m(n)|\Omegam|^{n}\Vc(\minus \Omegam^{-1})}
\label{eq:IL_Rayleigh1_Omega}
\end{equation}
where $\Omegam=\Thetam_1\Thetam_2^{-1}$, and $\omega_1,\ldots,\omega_m$ are the eigenvalues of
$\Omegam$. When $\Thetam_1$ and $\Thetam_2$ are scalar matrices, and
$\Omegam=\Thetam_1\Thetam_2^{-1}=\omega\Id$, the distribution of $\Lambdam$ can be obtained first by
applying the limit~\eqref{eq:limit_n0} to~\eqref{eq:IL_Rayleigh1_Omega}:
\begin{eqnarray}
p(\Lambdam) &=& \frac{\pi_m^2(p+n-m)!^m}{\Gamma_m(p)\Gamma_m(n)}
\frac{\Vc(\Lambdam)|\Lambdam|^{n-m}}{\omega^{mn}} \non
&&\quad \cdot \lim_{\Omegam\to
  \omega\Id}\frac{|\{{_1F_0}(p\plus n\minus m\plus 1;\,; -\lambda_j/\omega_i)\}|}{\Vc(-\Omegam^{-1})} \non
&=& \frac{\pi_m^2\Gamma_m(p\mathord{+}n)\Vc(\Lambdam)|\Lambdam|^{n-m}}{\Gamma_m(m)\Gamma_m(p)\Gamma_m(n)\omega^{mn}} \non
&& \quad \cdot|\{\lambda_j^{m-i}{_1F_0}(p\plus n\minus j\plus 1;\,; \minus \lambda_j/\omega)\}| 
\end{eqnarray}
and then by observing that
\begin{eqnarray}
& &\hspace{-6ex} |\{\lambda_j^{m-i} {_1F_0}(p+n-i+1;\,; -\lambda_j/\omega)\}| \non
&=& |\{\lambda_j^{m-i}(1+\lambda_j/\omega)^{-(p+n-i+1)}\}| \non
&=& |\{[\lambda_j/(1+\lambda_j/\omega)]^{m-i}(1+\lambda_j/\omega)^{-(p+n-m+1)}\}| \non
&=& \Vc(\Lambdam (\Id+\Lambdam/\omega))|\Id+\Lambdam/\omega|^{-(p+n-m+1)}  \non
&=& \Vc(\Lambdam)|\Id+\Lambdam/\omega|^{1-m}|\Id+\Lambdam/\omega|^{-(p+n-m+1)}  \non
&=& \Vc(\Lambdam)|\Id+\Lambdam/\omega|^{-(p+n)} \,.
\end{eqnarray}

\item For $m>n$, the distribution of the $n\times n$ random
  matrix $\Wm = \Hm_1\Herm(\Hm_2\Hm_2\Herm)^{-1}\Hm_1$ is given by~\cite[eq. (61)]{khatri}
\begin{equation}
p(\Wm) = \frac{\Gamma_{m}(n+p){_1F_0}(p\mathord{+}n; ; \Id\mathord{-}\Omegam^{-1}, \Wm(\Id\mathord{+}\Wm)^{-1})}{\Gamma_m(p)\Gamma_n(m)|\Omegam|^{n}|\Wm|^{n-m}|\Id+\Wm|^{p+n}}
\label{eq:pdfA_IL_Rayleigh2} 
\end{equation}
 where $\Omegam=\Thetam_1^{1/2}\Thetam_2^{-1}\Thetam_1^{1/2}$ is of size $m\times m$.  Note that,
 for any unitary matrix $\Vm$ independent of $\Wm$, we have $|\Vm\Wm\Vm\Herm|=|\Wm|$,
 $|\Id+\Vm\Wm\Vm\Herm|=|\Id+\Wm|$. Moreover, the eigenvalues of
 $\Vm\Wm\Vm\Herm(\Id+\Vm\Wm\Vm\Herm)^{-1}$ are the same as those of $\Wm(\Id+\Wm)^{-1}$. Thus,
 $p(\Vm\Wm\Vm)=p(\Wm)$. It follows that $\Wm$ is unitarily invariant.

Let $\Psim=\Wm(\Id+\Wm)^{-1}$ and the eigenvalues of $\Wm$ be
$\lambda_1,\ldots,\lambda_n$. Then, 
$\Psim$ has eigenvalues $\psi_j=\lambda_j/(1+\lambda_j)$, for $j=1,\dots,n$. In order to compute the
hypergeometric function of two matrix arguments of different size appearing
in~\eqref{eq:pdfA_IL_Rayleigh2}, we extend $\Psim$ to the $m\times m$ matrix $\widetilde{\Psim}$
given by
\[ \widetilde{\Psim} = \left[\begin{array}{cc} \Psim & \zerov \\ \zerov & \Em\end{array}\right] \]
where $\Em$ is an $(m-n)\times (m-n)$ matrix whose eigenvalues are $\ev=[e_1,\ldots,e_{m-n}]\Tran$. Then
the eigenvalues of $\widetilde{\Psim}$ are $\tilde{\psiv}
=[\psi_1,\ldots,\psi_n,e_1,\ldots,e_{m-n}]\Tran$.  It follows that
\begin{eqnarray}
&& \hspace{-6ex} {_1F_0}(p+n; ; \Id-\Omegam^{-1}, \Psim)  \non
&=& \lim_{\ev \to \zerov}{_1F_0}(p+n; ; \Id-\Omegam^{-1}, \widetilde{\Psim}) \non
&\stackrel{(a)}{=}& \lim_{\ev \to \zerov} \frac{\Gamma_m(m) (p+n-m)!^m}{\Gamma_m(p+n)} \frac{|\{f_{i}(\tilde{\psi}_j)\}|}{\Vc(\Id-\Omegam^{-1})\Vc(\widetilde{\Psim})} \non
&=& \frac{\Gamma_m(m) (p+n-m)!^m}{\Gamma_m(p+n)\Vc(\Id-\Omegam^{-1})} \lim_{\ev \to \zerov}  \frac{|\{f_{i}(\tilde{\psi}_j)\}|}{\Vc(\widetilde{\Psim})} \label{eq:hyper_2size}
\end{eqnarray}
where in $(a)$ we applied~\eqref{eq:pFq_2arguments} and $f_{i}(\tilde{\psi}_j)={_1F_0}(p+n-m+1; ;
(1-\omega_i^{-1})\tilde{\psi}_j)$, $i,j=1,\ldots,m$. By applying the limit in~\eqref{eq:limit} and
the properties in~\eqref{eq:determinant_property} and~\eqref{eq:hypergeometric_derivative}, the
limit in~\eqref{eq:hyper_2size} can be computed as
\begin{equation}
\lim_{\ev \to \zerov} \frac{|\{f_{i}(\tilde{\psi}_j)\}|}{\Vc(\widetilde{\Psim})} = \frac{\Gamma_m(p\plus n)\Gamma_n(m)}{\Gamma_n(p\plus n)\Gamma_m(m)}\frac{(p\plus n \minus m)!^{n-m}|\Fm|
  }{|\Psim|^{m-n}\Vc(\Psim)}
\label{eq:limit_IL_Rayleigh2}
\end{equation}
where for $i=1,\ldots,m$
\[ (\Fm)_{ij}\mathord{=}\left\{ \begin{array}{ll} f_{i}(\psi_j), & j=1,\ldots,n \\
 (1\minus \omega_i^{-1})^{m \minus j} & j=n\plus 1,\mathord{\ldots},m
  \end{array} \right.
\]
We now write the eigenvalue decomposition of $\Wm$ as $\Wm=\Um\Lambdam\Um\Herm$, where
$(\lambda_1,\ldots,\lambda_n)=\diag(\Lambdam)$. We then observe that $|\Wm| = |\Lambdam|$,
$|\Id+\Wm|=|\Id+\Lambdam|$, $|\Psim| = |\Lambdam||\Id+\Lambdam|^{-1}$, and that
$\Vc(\Psim)=\Vc(\Lambdam)|\Id+\Lambdam|^{1-n}$. Therefore, by
using~\eqref{eq:pdfA_IL_Rayleigh2},~\eqref{eq:hyper_2size}, and~\eqref{eq:limit_IL_Rayleigh2}, the
pdf of $\Wm$ can be rewritten as
\begin{equation}
p(\Wm) = \frac{\Gamma_m(p\plus n)(p\plus n\minus m)!^n}{\Gamma_n(p\plus n)\Gamma_m(p)}
\frac{|\Id\plus\Lambdam|^{m-p-n-1}|\Fm|}{|\Omegam|^{n}\Vc(\Id\minus\Omegam^{-1}) \Vc(\Lambdam)} \,.
\label{eq:pA_int_limited_Gaussian2}
\end{equation}
The pdf of the ordered eigenvalues of a complex random $n
\times n$ matrix $\Wm$ is given by~\cite[eq. (93)]{james}: 
\[  p(\Lambdam) = \frac{\pi_n^2\Vc(\Lambdam)^2}{\Gamma_n(n)}\int p_{\Wm}(\Um\Lambdam\Um\Herm)\dd \Um \,.\]
In our case, since $p(\Um\Lambdam\Um\Herm)$ does not depend on $\Um$, we
obtain~\eqref{eq:IL_Rayleigh2}.
\end{itemize}

\section{Proof of Lemma~\ref{def:non_central_2wishart}\label{app:IL_Rice1}}
\begin{itemize}
\item For $m\le n$, let us define $\Wm_2 = \Hm_2\Hm_2\Herm$. Then the matrix
  $\Wm=\Wm_2^{-1/2}\Hm_1\Hm_1\Herm \Wm_2^{-1/2}$ can be rewritten as $\Wm=\Hm\Hm\Herm$ where
  $\Hm=\Wm_2^{-1/2}\Hm_1$. For any given matrix $\Wm_2$, for $\Thetam=\theta\Id$ and
  $\Mm\Mm\Herm=\mu\Id$, $\Hm$ is a Gaussian complex matrix with average $\widetilde{\Mm}
  =\Wm_2^{-1/2}\Mm$ and independent columns whose covariance is $\Sigmam=\theta\Wm_2^{-1}$.  It
  follows that, given $\Wm_2$, $\Wm$ is a non central Wishart matrix
  and $p(\Wm |\Wm_2)$ is given
  by~\cite[eq. (99)]{james}
\begin{eqnarray}
 p_{\Wm |\Wm_2}(\Wm |\Wm_2) 
&=& {_0F_1}(;n; \Sigmam^{-1}\widetilde{\Mm}\widetilde{\Mm}\Herm\Sigmam^{-1}\Wm)\non
&&\quad \cdot \frac{\ee^{-\trace\{\Sigmam^{-1}\Wm\}}|\Wm|^{n-m}}{\ee^{\trace\{\Sigmam^{-1}\widetilde{\Mm}\widetilde{\Mm}\Herm\}}\Gamma_m(n)|\Sigmam|^n} \non
&=& {_0F_1}(;n; \theta^{-2}\mu \Wm_2\Wm) \non
&& \quad\cdot \frac{\ee^{-\trace\{\theta^{-1}\Wm_2\Wm\}}|\Wm_2|^n}{\ee^{\mu
    m/\theta}\theta^{nm}\Gamma_m(n)|\Wm|^{m-n}} \,. \nonumber
\end{eqnarray}
On the other hand, $\Wm_2$ is a central Wishart with covariance $\theta\Id$. Thus, the density of
$\Wm$ can be written as
\begin{eqnarray}
p_{\Wm}(\Wm) 
&=& \int p_{\Wm |\Wm_2}(\Wm |\Wm_2) p_{\Wm_2}(\Wm_2) \dd \Wm_2 \non
&=& \int {_0F_1}(;n;\frac{\mu}{ \theta^2} \Wm_2\Wm)\non
&&\quad\cdot \frac{\ee^{-\trace\{\Wm_2(\Id+\Wm)/\theta\}}|\Wm_2|^{p+n-m} \dd \Wm_2}{\ee^{\mu m/\theta}\theta^{(p+n)m} \Gamma_m(n)\Gamma_m(p)|\Wm|^{m-n}}  \non
&=& \frac{|\Wm|^{n-m}}{\ee^{\mu m/\theta}\theta^{(p+n)m} \Gamma_m(n)\Gamma_m(p)} \non
&& \quad \cdot \int \frac{{_0F_1}(;n; \frac{\mu}{\theta^2} \Wm_2\Wm)\dd \Wm_2 }{\ee^{\trace\{\Wm_2(\Id+\Wm)/\theta\}}|\Wm_2|^{m-p-n}}\,. \nonumber
\end{eqnarray}

In order to solve the above integral, we employ the following result
\[ \int\displaylimits_{\Bm=\Bm\Herm >0} \hspace{-2ex}
\frac{{_pF_q}(\av; \bv; \Cm\Bm)}{|\Bm|^{m-c}\ee^{\trace\{\Am\Bm\}}} \dd \Bm =
\frac{{_{p+1}F_q}(\av, c;\bv; \Cm\Am^{-1})}{[\Gamma_m(c)]^{-1}|\Am|^{c}} \,, \]
which holds for $m \times m$ matrices $\Am,\Bm$, and $\Cm$, and for
$\RR(c)>m-1$~\cite[eq. (115)]{McKayCollings}. Then, 
\[ p_{\Wm}(\Wm) =  \frac{\Gamma_m(p\plus n){_1F_1}(p\plus n;n;\frac{\mu}{\theta}\Wm(\Id\plus \Wm)^{-1})}{\Gamma_m(n)\Gamma_m(p)\ee^{\mu m/\theta}|\Wm|^{m-n}|\Id\plus \Wm|^{p+n}} \,. \] 
It can be observed that $p(\Wm)$ depends only on the eigenvalues of $\Wm$, thus it is unitarily
invariant. It follows that the pdf of the ordered eigenvalues of $\Wm$ is given by~\cite[eq. (93)]{james}
\[ p(\Lambdam) = \frac{\pi_m^2\Vc(\Lambdam)^2}{\Gamma_m(m)}\int p_{\Wm}(\Um\Lambdam\Um\Herm)\dd \Um \,,\]
which provides the results in~\eqref{eq:PL_IL_Rice1}.

\item For $m>n$, the distribution of the $n\times n$ matrix
  $\Wm=\Hm_1\Herm(\Hm_2\Hm_2\Herm)^{-1}\Hm_1$ is given by~\cite[eq. (105)]{james}
\begin{equation}
p(\Wm) = 
\frac{\Gamma_n(p\plus n){_1F_1}(p\plus n;m;\Omegam(\Id\plus \Wm^{\minus 1})^{\minus 1})}{\Gamma_n(m)\Gamma_n(p\plus n\minus m)\ee^{\trace\{\Omegam\}}|\Wm|^{n-m}|\Id\plus \Wm|^{p+n}}
\label{eq:PW_IL_Rice1}
\end{equation}
and the distribution of its eigenvalues is given by
\begin{equation}
p(\Lambdam) = \frac{p!^n}{(m\minus n)!^n}\frac{\pi_n \ee^{-\trace\{\Omegam\}}\Vc(\Lambdam)|\Fm||\Lambdam|^{m-n}}{\Gamma_n(p\plus n\minus m)\Vc(\Omegam)|\Id\plus \Lambdam|^{1+p}}
\label{eq:PL_IL_Rice2}
\end{equation}
where $\Omegam=\Mm\Herm\Thetam^{-1}\Mm$, $(\Fm)_{ij}={_1F_1}(p+1; m-n+1;\lambda_j\omega_i/(1+\lambda_j))$, and $\omega_1,\ldots,\omega_n$ are the eigenvalues of $\Omegam$. 
This result has been obtained by applying~\eqref{eq:pFq_2arguments} to~\cite[eq. (106)]{james}.

In the particular case where $\Omegam$ is a scalar matrix (i.e., $\Omegam=\omega\Id$),  matrix $\Wm$ is
unitarily invariant since its pdf in~\eqref{eq:PW_IL_Rice1} only depends on its eigenvalues
$\Lambdam$. Indeed, $|\Wm|=|\Lambdam|$, $|\Id+\Wm| = |\Id+\Lambdam|$, and the generalized
hypergeometric function ${_1F_1}(p+n;m;\omega(\Id+\Wm^{-1})^{-1})$ only depends on the eigenvalues
of its matrix argument, i.e., on $\Lambdam$. In such a case, the distribution of $\Lambdam$ can be
obtained from~\eqref{eq:PL_IL_Rice2} by applying the limit in~\eqref{eq:limit_n0} to the ratio
$|\Fm|/\Vc(\Omegam)$ and the property in~\eqref{eq:hypergeometric_derivative}. The result is reported
in~\eqref{eq:PL_IL_Rice2_omega}.
\end{itemize}

\section{Proof of Proposition \ref{prop:marginal}}\label{app.A}
The proof of~\eqref{eq:beta_eig2} and~\eqref{eq:beta_eig1} follows from the application
of~\cite[Theorem I]{icc06} to~\eqref{eq:beta2} and~\eqref{eq:beta1}, respectively.

The density given in~\eqref{eq:beta1} is an ordered eigenvalue distribution and the unordered
eigenvalue distribution is obtained by dividing~\eqref{eq:beta1} by $n!$.  Then, applying the
Laplace determinant expansion, the unordered eigenvalues distribution becomes
\begin{eqnarray}
p(\Lambdam)&=&\frac{\pi_n^2\Gamma_n(p+q)(1-\lambda_1)^{q-n}\lambda_1^{p-n-2}}{n!\Gamma_n(n)\Gamma_n(p)\Gamma_n(q)} \non
&& \cdot \sum_{i=1}^n\sum_{j=1}^n (-\lambda_1)^{i+j} \non
&&\quad \cdot\prod_{k=2}^n(1-\lambda_k)^{q-n}\lambda_k^{q-n}|\bar{\Vm}(\Lambdam)||\widetilde{\Vm}(\Lambdam)|
\non \label{eq:unorderedpdf}
\end{eqnarray}
where $\bar{\Vm}(\Lambdam)$ and $\widetilde{\Vm}(\Lambdam)$ are $(n-1)\times (n-1)$ matrices
obtained by deleting the first row and column from the Vandermonde matrix $\Vm(\Lambdam)$ and its
conjugate transpose, separately. The $(i,j)$-th entry of $\Vm(\Lambdam)$ and its conjugate transpose
are $\lambda_i^{j-1}$ and $\lambda_j^{i-1}$, respectively. Thanks to~\cite[Corollary 1]{win}, the
result in~\eqref{eq:beta_eig1} can be obtained through integration over $n-1$ eigenvalues from
$\lambda_2$ to $\lambda_n$.  The final expressions are in both cases due to the definition of the
scalar Beta function~\cite{abram}. It should be noticed that the choice of $\lambda_1$
in~\eqref{eq:unorderedpdf} has no effect on the final result, since we started from an unordered
eigenvalue distribution. Using the same approach, the proof of~\eqref{eq:beta_eig2} is
straightforward.

\section{Proof of~\eqref{eq:pY_pSigma1} and~\eqref{eq:pY_pSigma2}\label{app:main_result}}
We first observe that for $m > n$ the matrix $\Hm\Hm\Herm$ does not have full rank and has $m-n$
zero eigenvalues.  The $n$ non-zero eigenvalues of $\Hm\Hm\Herm$, denoted by
$\lambda_1,\ldots,\lambda_{n}$, are also the eigenvalues of $\Hm\Herm\Hm$ and are the elements of
the $n\times n$ diagonal matrix $\Lambdam$.  We start by rewriting~\cite[eq. (38)]{LozanoNihar} in
the case $m > n$ and obtain
\begin{equation}
p(\Ym) = \int \frac{\ee^{-\|\Ym\|^2}p(\Lambdam)}{\pi^{m b}|\Id\plus \gamma\Lambdam|^b}
 \left[\int  \ee^{\trace\{\widetilde{\Cm}\Um\Herm\Ym\Ym\Herm\Um\}} p(\Um|\Lambdam) \dd\Um \right]\!\!\dd\Lambdam
\label{eq:pY}
\end{equation}
 where $\Um$ is a unitary $m\times m$ matrix, $\widetilde{\Cm}$ is an $m\times m$ diagonal matrix
 whose elements are given by $(\widetilde{\Cm})_{jj}= c_j= \lambda_j\gamma/(1+\gamma\lambda_j)$,
 $j=1,\ldots, m$, with $c_j=0$, for $j=n+1,\ldots,m$.  Since we assume that $\Wm=\Hm\Herm\Hm$ is
 unitarily invariant, its eigenvalues do not depend on $\Um$. Moreover, $\Um$ is a Haar matrix (see
 Definition~\ref{def:unitarily_invariant}). Then, $p(\Um|\Lambdam)=p(\Um)$.  The inner integral over
 $\Um$ can be solved using the Harish-Chandra-Itzykson-Zuber integral~\cite{Itzykson-Zuber}
\[ \int_{\Uc(m)}  \ee^{\trace\{\widetilde{\Cm}\Um\Herm\Ym\Ym\Herm\Um\}}p(\Um) \dd \Um = \frac{\Gamma_m(m)|\Em|}{\pi_m\Vc(\widetilde{\Cm})\Vc(\Ym\Ym\Herm)}\,.  \]
The elements of  matrix $\Em$ are given by $(\Em)_{ij}=\ee^{y_ic_j}$, $i,j=1,\ldots, m$ and
$y_i$, $i=1,\ldots,m$, are the eigenvalues of $\Ym\Ym\Herm$.  Due to the fact that $c_j=0$ for
$j=n+1,\ldots,m$, we have $|\Em|=0$ and $\Vc(\widetilde{\Cm})=0$; thus the limit in~\eqref{eq:limit}
must be applied to the term $|\Em|/\Vc(\widetilde{\Cm})$. We have
\begin{equation}
\lim_{c_{n+1},\dots,c_{m} \to 0}\frac{|\Em|}{\Vc(\widetilde{\Cm})} =
\frac{\pi_m\Gamma_n(m)}{\pi_n\Gamma_m(m)}\frac{|\widetilde{\Em}|}{\Vc(\Cm) |\Cm|^{m-n}} 
\label{eq:limit2}
\end{equation}
where $\widetilde{\Em}$ is an $m\times m$  matrix whose elements are
given by $(\widetilde{\Em})_{ij} = \ee^{y_i c_j}$  for $1\le j\le n$, and $(\widetilde{\Em})_{ij} =
y^{j-n-1}_i$ for  $n+1 \le  j\le m$.   Also, $\Cm$ is  an $n\times  n$ diagonal
matrix  whose elements  are  $(\Cm)_{jj}= c_j=\lambda_j\gamma/(1+\gamma\lambda_j)$,
$j=1,\ldots, n$.  Therefore,~\eqref{eq:pY} can be rewritten as
\begin{equation}
p(\Ym) = \frac{\Gamma_{n}(m)\ky }{\pi_n} \int \frac{p(\Lambdam)|\widetilde{\Em}|}{|\Id+\gamma\Lambdam|^b}
\frac{|\Cm|^{n-m}}{\Vc(\Cm)} \dd\Lambdam 
\label{eq:pY2}
\end{equation}
where $\ky= \ee^{-\|\Ym\|^2}/(\Vc(\Ym\Ym\Herm)\pi^{m b})$ was defined in~\eqref{eq:ky}.
Since $c_j=\lambda_j\gamma/(1+\gamma\lambda_j)$, by applying the definition of the Vandermonde
determinant, we get $\Vc(\Cm) = |\Id+\gamma\Lambdam|^{1-n}\Vc(\gamma\Lambdam)$. Moreover, $|\Cm|
= |\gamma\Lambdam||\Id+\gamma\Lambdam|^{-1}$.  By substituting these
results in~\eqref{eq:pY2}, we obtain~\eqref{eq:pY_pSigma2}.  

\section{Proof of Proposition~\ref{GaussianRician}\label{app:fading_rice}}
We first observe that the matrix $\Hm$ in~\eqref{eq:ricean_channel} can be written as
$\Hm=\Hm_0/\sqrt{1+\kappa}$, where $\Hm_0=\sqrt{\kappa}\bar{\Hm}+\widetilde{\Hm}$.
\begin{itemize}
\item For $m \leq n$ and $\bar{\Hm}\bar{\Hm}\Herm=h\Id$, the joint distribution of the ordered
  eigenvalues of $\Hm_0\Hm_0\Herm$ is given by~\eqref{eq:non_central_wishart_mu} where $\mu=\kappa
  h$, i.e.,
\[
p_0(\Lambdam_0) =  \frac{\pi_m^2|\Lambdam_0|^n\Vc(\Lambdam_0) |\{\lambda_{0j}^{-i}{_0F_1}(\,; n\minus i\plus 1; \kappa h \lambda_{0j})\}| }{\Gamma_m(m)\Gamma_m(n)\ee^{\kappa h m+\trace\{\Lambdam_0\}}} 
\]
where $(\lambda_{01},\ldots,\lambda_{0m})=\diag(\Lambdam_0)$. Then, the pdf of the ordered eigenvalues of $\Hm\Hm\Herm$ is given by 
\begin{eqnarray}
p(\Lambdam) &=& (1+\kappa)^{m}p_0((1+\kappa)\Lambdam) \non
&=&\frac{\pi_m^2(1+\kappa)^{mn}|\Lambdam|^n |\Fm| \Vc(\Lambdam)}{\Gamma_m(m)\Gamma_m(n)\ee^{\kappa h m+(1+\kappa)\trace\{\Lambdam\}}}
\label{eq:PL_Rice}
\end{eqnarray}
where $(\Fm)_{ij} = \lambda_j^{-i}{_0F_1}(\,; n-i+1; \kappa(1+\kappa) h \lambda_j)$, $i,j=1,\ldots,
m$.  By substituting this equation in~\eqref{eq:pY_pSigma1} and by applying the result in
Appendix~\ref{app:integral_2determinants}, we obtain~\eqref{eq:pYRice1}.

\item For $m > n$, and for $\bar{\Hm}\Herm\bar{\Hm} = h\Id$, we adopt a procedure similar to the one
  above. In this case, the pdf of the non-zero eigenvalues of $\Hm\Hm\Herm$ is given
  by~\eqref{eq:PL_Rice} where $n$ and $m$ should be replaced by $m$ and $n$, respectively. By
  substituting $p(\Lambdam)$ in~\eqref{eq:pY_pSigma2} and by applying the result in
  Appendix~\ref{app:integral_2determinants}, we obtain~\eqref{eq:pYRice2}.
\end{itemize}

\section{Proof of Proposition~\ref{GaussianLMS}\label{app:LMS}}
For $m \leq n$, the distribution of the ordered eigenvalues of $\Hm\Hm\Herm$ is expressed
as~\cite[eq. (9)]{lmsmimo}

\begin{equation}
\label{eq:pSigma_LMS1}
p(\Lambdam) 
=  \frac{\Gamma(\alpha\minus m\plus 1)^m}{\Gamma(n\minus m\plus 1)^m}\frac{\pi_m\ee^{-\trace\{\Lambdam\}}}{\Vc((\Id\plus \Omegam)^{-1}) }  \frac{\Vc(\Lambdam) |\Lambdam|^{n-m}|\Fm|}{\Gamma_{m}(\alpha)|\Id\plus \Omegam^{-1}|^\alpha}
\end{equation}
with $(\Fm)_{ij}= {_1F_1}(\alpha-m+1;n-m+1; \lambda_j/(1+\omega_i))$. 
When $\Omegam=\omega\Id$,  the
expression of $p(\Lambdam)$ can be derived from~\eqref{eq:pSigma_LMS1} by applying the limit
in~\eqref{eq:limit_n0} and by using the property in~\eqref{eq:determinant_property}. For simplicity, we define $\Thetam = (\Id+\Omegam)^{-1} = \theta\Id$ where
$\theta=(1+\omega)^{-1}$.  Then, 
\begin{eqnarray}
\label{eq:pSigma_LMS1_limit}
p(\Lambdam) 
&=& \frac{\pi_m\Gamma(\alpha\minus m\plus 1)^m \Vc(\Lambdam) |\Lambdam|^{n-m}}{\Gamma_{m}(\alpha)\Gamma(n\minus m\plus 1)^m\ee^{\trace\{\Lambdam\}}(1\plus 1/\omega)^{m\alpha}} \non
&&\qquad \cdot \lim_{\Thetam\to \theta\Id }\frac{|\Fm|}{\Vc(\Thetam)} \non
&=&  \frac{\pi_m^2\Vc(\Lambdam) |\Lambdam|^n|\widetilde{\Fm}|}{\Gamma_m(m)\Gamma_{m}(n)\ee^{\trace\{\Lambdam\}}(1+1/\omega)^{m\alpha}} \non
\end{eqnarray}
where $(\widetilde{\Fm})_{ij} =\lambda_j^{-i} {_1F_1}(\alpha-i+1;n-i+1; \lambda_j/(1+\omega))$,
$i,j=1,\ldots,m$.  The proposition statement follows by replacing~\eqref{eq:pSigma_LMS1_limit}
in~\eqref{eq:pY_pSigma1}. Similarly, when $m > n$ and $\Omegam=\omega\Id$, the
  distribution of the eigenvalues of $\Hm\Herm\Hm$ is given by
\[ p(\Lambdam) = \frac{\pi_n^2\ee^{-\trace\{\Lambdam\}}\Vc(\Lambdam) |\Lambdam|^m}{\Gamma_n(m)(1+1/\omega)^{n\alpha}}\frac{|\widetilde{\Fm}|}{\Gamma_n(n)} \]
where $(\widetilde{\Fm})_{ij} =\lambda_j^{-i} {_1F_1}(\alpha-i+1;m-i+1; \lambda_j/(1+\omega))$,
$i,j=1,\ldots,n$.  Again, the proposition statement is obtained by replacing the above equation
in~\eqref{eq:pY_pSigma2}.

\section{Proof of Proposition~\ref{RiceIL}\label{app:PY_IL_Rice}}
 We first observe that the matrix $\Hm$ in~\eqref{eq:interference_limited} can be written as
 $\Hm=\Hm_0/\sqrt{1+\kappa}$, where
\[ \Hm_0 =  \sqrt{\kappa}\left(\widehat{\Hm}\widehat{\Hm}\Herm\right)^{-1/2}\bar{\Hm}_s+\left(\widehat{\Hm}\widehat{\Hm}\Herm\right)^{-1/2}\widetilde{\Hm}_s \]
 
\begin{itemize}
\item for $m\leq n$, and $\bar{\Hm}_s\bar{\Hm}_s\Herm=h\Id$, the distribution of the ordered
  eigenvalues of $\Hm_0\Hm_0\Herm$ is given by~\eqref{eq:PL_IL_Rice1} 
\begin{eqnarray}
 p_0(\Lambdam_0) &=& \frac{\pi_m^2\ee^{-h\kappa m/\theta}\Gamma_m(Ln\plus n)\Vc(\Lambdam_0)^2|\Lambdam_0|^{n-m}}{\Gamma_m(m)\Gamma_m(n)\Gamma_m(Ln)|\Id+\Lambdam_0|^{Ln+n}} \non
 && \qquad \cdot {_1F_1}\left(Ln\plus n;n;\frac{h\kappa}{\theta}\Lambdam_0(\Id+\Lambdam_0)^{-1}\right) \nonumber
\end{eqnarray}
where we set $\mu=h\kappa$.
The distribution of the eigenvalues of $\Hm\Hm\Herm$ can then be obtained as
\begin{eqnarray}
 p(\Lambdam) 
&=& \tilde{\kappa}^m p_0(\tilde{\kappa}\Lambdam) \non
&=& \frac{\pi_m^2\tilde{\kappa}^{mn}\Gamma_m(Ln\plus n)|\Lambdam|^{n-m}\Vc^2(\Lambdam)}{\Gamma_m(m)\Gamma_m(n)\Gamma_m(Ln)\ee^{h\kappa
    m/\theta}|\Id\plus \tilde{\kappa}\Lambdam|^{Ln+n}}\non
&&\quad \cdot\, {_1F_1}\left(Ln\plus n;n;\frac{h\kappa\tilde{\kappa}}{\theta}\Lambdam(\Id\plus \tilde{\kappa}\Lambdam)^{-1}\right) \non
\end{eqnarray}
where $\tilde{\kappa}=1+\kappa$.
By substituting the above expression in~\eqref{eq:pY_pSigma1}, 
and by exploiting the property~\cite[eq. (2.36)]{McKay}
\[ {_1F_1}(a;b;\Psim) = \frac{|\{{_1F_1}(a\minus m\plus j;b\minus m\plus j;\psi_i)\psi_i^{j-1}\}|}{\Vc(\Psim)}\,,\]
which holds for any $m\times m$ Hermitian matrix $\Psim$ with eigenvalues $\psi_1,\ldots,\psi_m$,
we obtain~\eqref{eq:PY_IL_Rice1}.

\item For $m>n$, the distribution of the eigenvalues of $\Hm_0\Hm_0\Herm$ is given
  by~\eqref{eq:PL_IL_Rice2_omega}: 
\[ p_0(\Lambdam_0) = \frac{\pi_n^2\Gamma_n(Ln+n)\ee^{-\omega n}|\Fm||\Lambdam_0|^{m-n}\Vc(\Lambdam_0)}{\Gamma_n(n)\Gamma_n(m)\Gamma_n(Ln+n-m)|\Id+\Lambdam_0|^{Ln+1}} \]
 where $\Omegam=\omega\Id=\Mm\Herm\Thetam^{-1}\Mm$. In our case we have
  $\Mm = \sqrt{\kappa}\bar{\Hm}_s$, thus $\omega\Id=\kappa\bar{\Hm}_s\Herm\Thetam^{-1}\bar{\Hm}_s$. 
It follows that the matrix $\Hm\Hm\Herm$ is unitarily invariant if $\bar{\Hm}_s\Herm\Thetam^{-1}\bar{\Hm}_s=\omega/\kappa \Id$. The distribution of the eigenvalues of $\Hm\Hm\Herm$ can then be obtained as
\begin{eqnarray} p(\Lambdam) 
&=& \tilde{\kappa}^n p_0(\tilde{\kappa}\Lambdam) \non
&=& \frac{\pi_n^2\Gamma_n(Ln+n)|\Lambdam|^{m-n}\Vc(\Lambdam)}{\Gamma_n(n)\Gamma_n(m)\Gamma_n(Ln+n-m)} \non
&&\cdot 
\frac{|\{\tilde{\lambda}_j^{n-i}{_1F_1}(Ln\plus n\minus i\plus 1;m\minus i\plus 1;\omega\tilde{\kappa}\tilde{\lambda}_j)\}|}{\tilde{\kappa}^{-nm}\ee^{\omega n}|\Id+\tilde{\kappa}\Lambdam|^{Ln+1}} \nonumber
\end{eqnarray} 
where $\tilde{\kappa}=1+\kappa$ and
$\tilde{\lambda}_j=\lambda_j/(1+\tilde{\kappa}\lambda_j)$. By substituting this expression in~\eqref{eq:pY_pSigma2}, we obtain~\eqref{eq:PY_IL_Rice2}.
\end{itemize}

\section{Proof of Proposition~\ref{uncond}\label{app:uncond}}
Given the above assumptions and considering that
$\Xm=\sqrt{c}\Dm^{1/2}\Phim$, $p(\Ym|\Dm)$ is given in as~\cite[eq.~(53)]{durisi}: 
\begin{eqnarray}    
p(\Ym|\Dm) 
&=& \frac{\ee^{-\|\Ym\|^2}A}{\pi^{mb}|\Id\plus \gamma c\Dm|^m} 
\end{eqnarray}
where 
\begin{eqnarray} 
A &=& \int\displaylimits_{\Sc(b,n)}\ee^{\trace\{\Deltam\Phim\Ym\Herm\Ym\Phim\Herm\}}p(\Phim)\dd \Phim \non
 &=& \frac{1}{|\Sc(b,n)|} \int\displaylimits_{\Sc(b,n)}\ee^{\trace\{\Deltam\Phim\Ym\Herm\Ym\Phim\Herm\}}\dd
\Phim  \nonumber
\end{eqnarray}
and $\Deltam = \gamma c \Dm(\Id+\gamma c \Dm)^{-1}$.
In~\cite[Appendix A]{durisi}, it is observed that the integral above is not an instance
of the Harish-Chandra-Itzykson-Zuber (HCIZ) integral~\cite{Itzykson-Zuber} since the 
  $n\times b$ matrix $\Phim$ is not a square matrix. 
In order to circumvent this problem, one has to extend matrix $\Phim\Herm$ to the unitary $b \times b$ Haar matrix $\widetilde{\Phim}\Herm= [\Phim\Herm,
  \Phim_{\perp}\Herm]$, where $\Phim_{\perp}\Herm$ is the orthogonal complement of $\Phim\Herm$ with
respect to the unitary group $\Uc(b)$. Thus, following~\cite[Appendix A]{durisi}, we can write
\[ A = 
\frac{1}{|\Sc(b,n)||\Uc(b\minus
  n)|}\int\displaylimits_{\Uc(b)}\ee^{\trace\{\Deltam\Phim\Ym\Herm\Ym\Phim\Herm\}} \dd
\widetilde{\Phim} \,. \] Next, the $n \times n$ diagonal matrix
$\Deltam=\diag(\delta_1,\ldots,\delta_n)$ can be extended to the $b\times b$ matrix
$\widetilde{\Deltam}=\diag(\delta_1,\ldots,\delta_n,q_1,\ldots,q_{b-n})$ where the elements of
$\qv=[q_1,\ldots,q_{b-n}]$ are distinct and different from $\delta_1,\ldots,\delta_n$. The above
integral can then be written as
\begin{equation}
A = \frac{1}{|\Sc(b,n)||\Uc(b\minus
  n)|}\lim_{\qv \to 0}\int\displaylimits_{\Uc(b)}
\ee^{\trace\{\widetilde{\Deltam}\widetilde{\Phim}\Ym\Herm\Ym\widetilde{\Phim}\Herm\}}
\dd \widetilde{\Phim}  \,.
\label{eq:Delta_extension}
\end{equation}
We  observe that the matrix $\Ym\Herm\Ym$  has $b-m$ zero-eigenvalues, and its
non-zero eigenvalues are the eigenvalues of $\Ym\Ym\Herm$.  Since the HCIZ integral is a function of
the eigenvalues of the matrices $\widetilde{\Deltam}$ and $\Ym\Herm\Ym$, we replace the matrix
$\Ym\Herm\Ym$ with the $b\times b$ block diagonal matrix $\Psim= \diag(\Ym\Ym\Herm, \Pm)$ where
$\Pm$ is diagonal and has diagonal entries $\pv=[p_1,\ldots, p_{b-m}]$. Such elements are positive,
distinct, and they are different from the eigenvalues of
$\Ym\Ym\Herm$.  In conclusion, we can write: 
\begin{eqnarray}
A &=& \frac{1}{|\Sc(b,n)||\Uc(b-n)|} \lim_{\qv \to \zerov}\lim_{\pv \to \zerov} \int\displaylimits_{\Uc(b)} \ee^{\trace\{\widetilde{\Deltam}\widetilde{\Phim}\Psim\widetilde{\Phim}\Herm\}} \dd \widetilde{\Phim} \non
&=&  \frac{\Gamma_b(b)|\Uc(b)|}{\pi_b|\Sc(b,n)||\Uc(b-n)|}\lim_{\qv \to 0}\lim_{\pv \to \zerov} \frac{|\Fm|}{\Vc(\Psim)\Vc(\widetilde{\Deltam})} \non
&=&  \frac{\Gamma_b(b)}{\pi_b}\lim_{\qv \to 0}\lim_{\pv \to \zerov} \frac{|\Fm|}{\Vc(\Psim)\Vc(\widetilde{\Deltam})}
\label{eq:HCIZ_integral}
\end{eqnarray}
where $(\Fm)_{ij} = \ee^{\psi_i\tilde{\delta}_j}$ and $\psi_i$
and $\tilde{\delta}_j$ are the eigenvalues of $\Psim$ and $\widetilde{\Deltam}$, respectively.
In~\eqref{eq:HCIZ_integral} we first used the HCIZ integral~\cite{Itzykson-Zuber}
and then the equality $|\Uc(b)|= |\Sc(b,n)||\Uc(b-n)|$.
Then, we apply twice the limit in~\eqref{eq:limit} 
and obtain: 
\begin{eqnarray}
\lim_{\qv \to 0}\lim_{\pv \to \zerov} \frac{|\Fm|}{\Vc(\Psim)\Vc(\widetilde{\Deltam})} &=&
\frac{\pi_b\Gamma_m(b)|\Ym\Ym\Herm|^{m-b}}{\pi_m\Gamma_b(b)\Vc(\Ym\Ym\Herm)} \non
&&\quad \cdot \frac{\pi_b\Gamma_n(b)|\widehat{\Fm}||\Deltam|^{n-b}}{\pi_n\Gamma_b(b)\Vc(\Deltam)} \label{eq:two_limits}
\end{eqnarray}
where  for $m \leq n$,
\begin{equation}
  (\widehat{\Fm})_{ij} = \left\{
  \begin{array}{ll}
    \ee^{y_i\delta_j} & i=1,\ldots,m; j=1,\ldots,n \\
    y_i^{b-j}  &  i=1,\ldots,m; j=n+1,\ldots,b \\
    \delta_j^{b-i}  & i=m+1,\ldots,b, j=1,\ldots,n \\
    (b-i)!   & i=j; j= n+1,\ldots,b \\
    0 & \mbox{elsewhere} \,,
  \end{array}
  \right.
 \label{eq:F_definition1}
\end{equation}
while for $m > n$,
\begin{equation}
  (\widehat{\Fm})_{ij} = \left\{
  \begin{array}{ll}
    \ee^{y_i\delta_j} & i=1,\ldots,m; j=1,\ldots,n \\
    y_i^{b-j}  &  i=1,\ldots,m; j=n+1,\ldots,b \\
    \delta_j^{b-i}  & i=m+1,\ldots,b, j=1,\ldots,n \\
    (b-i)!   & i=j; j= m+1,\ldots,b \\
    0 & \mbox{elsewhere.}
  \end{array}
  \right.
 \label{eq:F_definition2}
\end{equation}
In summary, 
\begin{equation}
p(\Ym|\Dm) 
= \frac{\pi_b\Gamma_m(b)\Gamma_n(b)\ky  |\Ym\Ym\Herm|^{m-b}|\widehat{\Fm}||\Deltam|^{n-b}}{\pi_m\pi_n\Gamma_b(b)\Vc(\Deltam)|\Id\plus \gamma c\Dm|^m}
\label{eq:PY|D_4}
\end{equation}
where $\ky$ was defined in~\eqref{eq:ky}.

We now focus on the case $m > n$ and compute the determinant $|\widehat{\Fm}|$.
Note that $\widehat{\Fm}$ can be written as
\[ \widehat{\Fm} = \left[\begin{array}{cc}\widehat{\Fm}_1 & \widehat{\Fm}_2 \\ \widehat{\Fm}_3 & \widehat{\Fm}_4  \end{array}  \right] \]
where $\widehat{\Fm}_1$ is of size $m\times m$, $\widehat{\Fm}_2$
$m\times (b-m)$, $\widehat{\Fm}_3$  $(b-m)\times m$, and
$\widehat{\Fm}_4$ $(b-m)\times (b-m)$. By using the property of the
determinant of block matrices~\cite{horn:85}, we have: 
\[ |\widehat{\Fm}| = |\widehat{\Fm}_4||\widehat{\Tm}| \,,\]
where $\widehat{\Tm} = \widehat{\Fm}_1-\widehat{\Fm}_2\widehat{\Fm}_4^{-1}\widehat{\Fm}_3$.
In our case, $\widehat{\Fm}_4$ is diagonal (see the definition of $\widehat{\Fm}$ in~\eqref{eq:F_definition2})
and $|\widehat{\Fm}_4| = \prod_{i=0}^{b-m-1}i!$. Moreover, we have $(\widehat{\Fm}_2\widehat{\Fm}_4^{-1}\widehat{\Fm}_3)_{ij} = \sum_{k=0}^{b-m-1}
(y_i\delta_j)^k/k!$ for $i=1,\ldots,m$, $j=1,\ldots,n$, and $(\widehat{\Fm}_2\widehat{\Fm}_4^{-1}\widehat{\Fm}_3)_{ij}=0$
otherwise.
It follows that for $i=1,\ldots,m$
\[
(\widehat{\Tm})_{ij} = \left\{
\begin{array}{ll}
\ee^{y_i\delta_j}-\sum_{k=0}^{b-m-1} \frac{(y_i\delta_j)^k}{k!} &  j=1,\ldots,n \\
y_i^{b-j}    & j=n+1,\ldots,m \,.
\end{array}
\right.
\]
Note that, for $i=1,\ldots,m$ and $j=1,\ldots,n$,
\begin{eqnarray}
(\widehat{\Tm})_{ij}
&=& \ee^{y_i\delta_j}-\sum_{k=0}^{b-m-1} \frac{(y_i\delta_j)^k}{k!} \non
&=& \sum_{k=0}^{\infty}\frac{(y_i\delta_j)^k}{k!}-\sum_{k=0}^{b-m-1} \frac{(y_i\delta_j)^k}{k!} \non
&=& \sum_{k=b-m}^{\infty}\frac{(y_i\delta_j)^k}{k!}\non
&=& \sum_{h=0}^{\infty}\frac{(y_i\delta_j)^{b-m+h}}{(b\minus m\plus h)!}\non
&=& (y_i\delta_j)^{b-m}\sum_{h=0}^{\infty}\frac{(y_i\delta_j)^h h!}{(b-m+h)! h!}\non
&=& \frac{(y_i\delta_j)^{b-m}}{(b\minus m)!}\sum_{h=0}^{\infty}\frac{(1)_h(y_i\delta_j)^h}{(b\minus m\plus 1)_hh!}\non
&=& \frac{(y_i\delta_j)^{b-m}}{(b\minus m)!}{_1F_1}(1; b\minus m \plus 1; y_i\delta_j)\nonumber
\end{eqnarray}
since $h! = (1)_h$ and $(b-m+h)! = (b-m+1)_h (b-m)!$.
Also, for $i=1,\ldots,m$ and $j=n+1,\ldots,m$, $(\widehat{\Tm})_{ij} = y_i^{b-j}= y_i^{b-m}y_i^{m-j}$.

As a consequence, the matrix $\widehat{\Tm}$
can be rewritten as $\widehat{\Tm} =
\Lm\Gm\Rm$, where $\Lm$ and $\Rm$ are diagonal $m \times m$ matrices given by, respectively,
$\Lm=\diag(y_1^{b-m},\ldots, y_m^{b-m})$, and
\[ \Rm=\diag(\delta_1^{b-m}/(b-m)!, \ldots,
\delta_n^{b-m}/(b-m)!,1,\ldots,1)\,.\] Furthermore, $\Gm$ is an $m \times m$ matrix whose elements,
for $i\mathord{=}1,\ldots,m$, are given by
\[ 
(\Gm)_{ij} = \left\{
\begin{array}{ll}
{_1F_1}(1; b- m+ 1; y_i\delta_j) & j=1,\ldots,n \\
y_i^{m-j}    & j=n+1,\ldots,m \,. 
\end{array}
\right.
\]
Thus, we have:
\begin{eqnarray}
|\widehat{\Fm}| 
&=& |\widehat{\Fm}_4||\widehat{\Tm}| \non
&=&  |\Lm||\Gm||\Rm| \prod_{i=0}^{b-m-1}i! \non
&=& \frac{|\Ym\Ym\Herm|^{b-m}|\Gm||\Deltam|^{b-m}}{(b-m)!^n}\prod_{i=0}^{b-m-1}i!\,.
\label{eq:|hatF|}
\end{eqnarray}
In conclusion, by substituting ~\eqref{eq:|hatF|} in~\eqref{eq:PY|D_4}, we
get~\eqref{eq:PY|D2}.

For $m \leq n$, a similar procedure can be used to compute the determinant $|\widehat{\Fm}|$. In
this case,
\[ |\widehat{\Fm}| = \frac{|\Ym\Ym\Herm|^{b-n}|\Gm||\Deltam|^{b-n}}{(b-n)!^m} \prod_{i=0}^{b-n-1}i!\,. \]
Again, by substituting the above expression in~\eqref{eq:PY|D_4}, we get~\eqref{eq:PY|D1}.
Here, however, the expression of $(\Gm)_{ij}$ changes as follows: 
\[ 
(\Gm)_{ij} = \left\{
\begin{array}{ll}
{_1F_1}(1; b-n+1; y_i\delta_j) & i=1,\ldots,m \\
\delta_j^{n-i}    & i=m+1,\ldots,n
\end{array}
\right.
\]
and for $j=1,\ldots,n$.

\section{Proof of Proposition~\ref{unconditional2}\label{app:unconditional2}}

The law of the output of a channel, as the one in (\ref{lin-model}), conditioned on the input power
allocation $\Dm$, is reported in \cite[eq. (58)]{durisi}, i.e.,
\begin{equation}\label{durisicond}
p(\Ym|\Dm)
= \frac{\Gamma_n(b)\ky|\Id+c\gamma\Dm|^{b-m-1}|\Fm|}{\pi_n\gamma_n|\gamma c \Dm|^{b-n}\Vc(c\Dm)}\,,
\end{equation}
where $\ky = \ee^{-\|\Ym\|^2}/(\pi^{mb}\Vc(\Ym\Ym\Herm))$ and $(\Fm)_{ij}=\exp\left(\frac{c\gamma y_i d_j}{1+c\gamma d_j}\right)$,
$i=1,\ldots,n,j=1,\ldots,b$, and $\Fm_{ij}=y^{b-j}_i$, $i=n+1,\ldots,b,j=1,\ldots,b$.

In order to take average of~\eqref{durisicond}, we first write
$|\Fm|$ as the product of two determinants. Indeed, we partition $\Fm$ as
\begin{equation}
\Fm=\left[ \begin{array}{cc}
\Fm_1 & \Fm_2\\
\Fm_3 & \Fm_4
\end{array} \right]
\label{eq:blockM}
\end{equation}
where $(\Fm_4)_{ij}=y_{n+i}^{b-n-j}\,i,j=1,\dots, b-n$, and $\Fm_1$ is the principle $n\times n$
submatrix of $\Fm$.  Applying the property of the determinant of block
matrices~\cite{horn:85} 
to~\eqref{eq:blockM}, we obtain
\begin{equation}\label{detM}
|\Fm|=|\Fm_4| |\Tm|\,,
\end{equation}
where $\Tm = \Fm_1-\Fm_2\Fm_4^{-1}\Fm_3$.  We notice that $|\Fm_4|$ is independent of $\Dm$, and the
matrix $\Tm$ has the same size as $\Dm$.  For $m > n$, $p(\Dm)$ is given by~\eqref{eq:beta1}. We
then get
\begin{eqnarray}\label{eq:P_Y1_massimve_MIMO}
p(\Ym)
&=& \int p(\Ym | \Dm) p(\Dm)\dd \Dm \non
&=& \frac{\Gamma_n(b)\ky |\Fm_4|}{\pi_n\gamma_n}
\int\frac{|\Id\plus c\gamma\Dm|^{b-m-1}|\Tm|p(\Dm)}{|\gamma c \Dm|^{b-n}\Vc(c\Dm)}\dd \Dm \non
&=&
\frac{\pi_n\Gamma_n(b)\Gamma_n(m)(\gamma c)^{n(n-b)}\ky |\Fm_4|}{\gamma_n\Gamma_n(n)c^{n(n-1)/2}\Gamma_n(b\minus n)\Gamma_n(n\plus m\minus b)} \non
&&\quad \cdot\int\frac{|\Id+c\gamma\Dm|^{b-m-1}|\Tm|}{|\Id-\Dm|^{b-m}|\Dm|^n}
\Vc(\Dm)
\dd \Dm\,, \non
&=&
\frac{\pi_n\Gamma_n(b)\Gamma_n(m)(\gamma c)^{n(n-b)}\ky |\Fm_4||\Zm|}{\gamma_nc^{n(n-1)/2}\Gamma_n(n)\Gamma_n(b\minus n)\Gamma_n(n\plus m\minus b)} 
\end{eqnarray}
where
\begin{eqnarray} 
(\Zm)_{ij}
&=&\int_0^1 \frac{(1+c\gamma x)^{b-m-1}x^{i-1-n}}{(1-x)^{b-m}} \non
&&\quad \cdot \left[\exp\left(\frac{c\gamma y_i x}{1+c\gamma x}\right)-(\Fm_2\Fm_4^{-1}\Fm_3)_{ij}\right] \nonumber
\end{eqnarray}
has been obtained by using the result in Appendix~\ref{app:integral_2determinants}, and
\[ (\Fm_2\Fm_4^{-1}\Fm_3)_{ij} = \sum_{\ell, k=1}^{b-n} (\Fm_4^{-1})_{\ell k}\exp\left(\frac{c\gamma x y_{\ell+n}}{1+c\gamma x}\right)y_j^{b-k-n}\,. \]

\section{Proof of Proposition~\ref{cond}\label{app:cond}}
Conditioned on the input $\Xm$, the output $\Ym$ is
complex Gaussian and has i.i.d. rows, so that the evaluation of the differential entropy
can be carried out by considering just an arbitrary row, $\yv$, of
$\Ym$ and, then, scaling the result by $m$.
We note that $\yv$ is multivariate Gaussian distributed  with
covariance equal to $(\Id+\gamma\Xm\Herm\Xm$). Thus, considering the
optimal input matrix, $\Xm=\sqrt{c}\Dm^{1/2}\Phim$ and  conditioning
on it, the differential entropy is given by: 
\begin{equation}\label{eq:prop3}
h(\yv|\Xm)=b\log_2(\pi e)+n \EE\left[\log_2\left(1+c\gamma\delta\right)\right]\,,
\end{equation}
with $\delta$ being distributed as a single unordered eigenvalue of
the matrix $\Dm$.
By using (\ref{eq:beta1}) and
considering $p=b-n$ and $q=m+n-b$, 
$p(\Dm)$ reads as 
\begin{eqnarray}
p(\Dm)= \frac{\pi_n^2\Gamma_{n}(m)|\Id-\Dm|^{m-b}|\Dm|^{b-2n}\Vc^2(\Dm)}{\Gamma_{n}(n)\Gamma_{n}(b-n)\Gamma_{n}(m+n-b)} \,. 
\end{eqnarray}
By exploiting the result given in Proposition~\ref{prop:marginal} and by denoting the constant terms
in the above expression by $K$, we obtain:
\begin{eqnarray}
p(\delta)=\frac{K}{n}\sum_{i,j=1}^{n} \delta^{(b-2n+i+j-2)}(1-\delta)^{(m-b)}a_{ij}
\end{eqnarray}
with $a_{ij}$ being defined as in the above proposition.
The integral in (\ref{eq:prop3}) can be solved by resorting to partial
integration. Indeed, taking $\log_2\left(1+c\gamma\delta\right)$ as
the primitive factor and recalling that
$(1-\delta)^{n-m}=\sum_{\ell=0}^{n-m}{n-m \choose \ell}(-1)^\ell \delta^\ell$,
by virtue of \cite[3.194.1]{Gradshteyn}, we obtain
\begin{eqnarray}\label{condoptintermclosed}
&& \EE\left[\log_2\left(1+\gamma\delta\right)\right]=\frac{K}{n}\sum_{i,j=1}^{n}a_{ij}
\sum_{\ell=0}^{m-b}{m-b \choose \ell}\frac{(-1)^\ell}{s_{i,j,\ell}-1}\nonumber\\
&&\cdot \left[\log_2(1+c\gamma)-c\gamma\frac{{_2F_1}\left(1, s_{i,j,\ell};s_{i,j,\ell}+1;-\gamma\right)}{\ln(2)(s_{i,j,\ell})} \right]\nonumber 
\end{eqnarray}
where $s_{i,j,\ell}= b-2n+i+j+\ell$.
Then, using this expression in~\eqref{eq:prop3}, we get~\eqref{conditional}.

\section{Lemma 2 in~\cite{Shin2006}\label{app:integral_2determinants}}
Consider a function $\xi(x)$, an arbitrary $n \times n$ matrix $\Phim(\xv)$ such that $(\Phim)_{ij}
= \phi_i(x_j)$, and an arbitrary $m \times m$ matrix $\Psim$, $m \ge n$, whose elements are given by
\[ 
(\Psim)_{ij} = \left\{    
\begin{array}{ll}
\psi_i(x_j) & 1\le i \le m, 1\leq j \leq n \\ 
c_{ij}  & 1\le i \le m, n+1\leq j \leq m 
\end{array}
\right.
\]
where $c_{ij}$ are constant. Then, the following identity holds:
\begin{equation}
 \int_{[a,b]^n} |\Phim(\xv)| |\Psim(\xv)| \prod_{k=1}^{n}\xi(x_k) \dd\xv = n! |\Xim|
 \label{eq:integral_2determinants}
\end{equation}
where, for $1\le i \le m$,
\[ 
(\Xim)_{ij} = \left\{    
\begin{array}{ll}
\int_a^b \psi_i(x) \phi_j(x)\xi(x)\dd x &  1\leq j \leq n \\ 
c_{ij}  &  n+1\leq j \leq m \,.
\end{array}
\right.
\]
For the specific case $m=n$, this result appears in~\cite[Corollary II]{win}.

\begin{IEEEbiography}[{\includegraphics[width=1in, height=1.25in, clip, keepaspectratio]{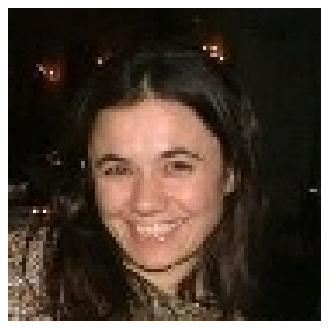}}]{Giusi Alfano}
received the Laurea degree in communication engineering from the University Federico II, Naples,
Italy, in 2001, where she was CNIT junior researcher from 2003 to 2005, and the phd in Information
Engineering from University of Sannio, Benevento, Italy, in 2007. She was visiting researcher at
ftw, Wien, in 2007 and at Chalmers University, Goteborg, in 2012. She was post-doc in 2009 in the
Alcatel-Lucent Chair of Flexible Communications, Sup\'elec, Paris, and in 2011 she was ERCIM
post-doc in NTNU, Trondheim, Norway.  She currently holds a post-doctoral position at Politecnico di
Torino, Italy.  Her research work lies mainly in the field of random matrix theory applications to
MIMO wireless communications and sensor networks.
\end{IEEEbiography}

\begin{IEEEbiography}[{\includegraphics[width=1in,height=1.25in,clip,keepaspectratio]{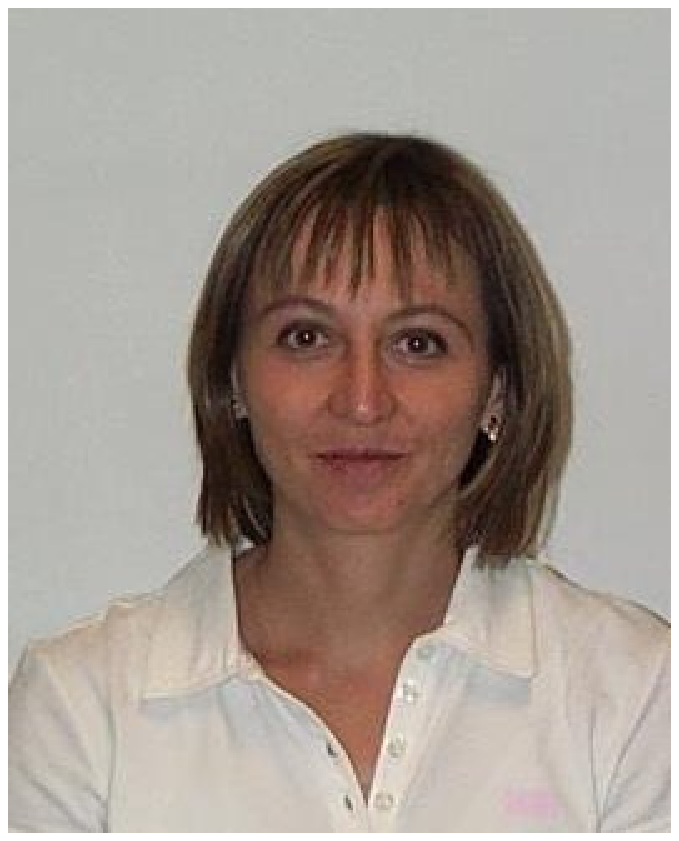}}]%
{Carla-Fabiana Chiasserini} (M'98, SM'09) graduated in Electrical Engineering (summa cum laude) from
the University of Florence in 1996. She received her Ph.D. from Politecnico di Torino, Italy, in
2000.  She has worked as a visiting researcher at UCSD in 1998--2003, and she is currently an
Associate Professor with the Department of Electronic Engineering and Telecommunications at
Politecnico di Torino.  Her research interests include architectures, protocols, and performance
analysis of wireless networks. Dr. Chiasserini has published over 200 papers in prestigious journals
and leading international conferences, and she serves as Associated Editor of several journals.
\end{IEEEbiography}

\begin{IEEEbiography}[{\includegraphics[width=1in,height=1.25in,clip,keepaspectratio]{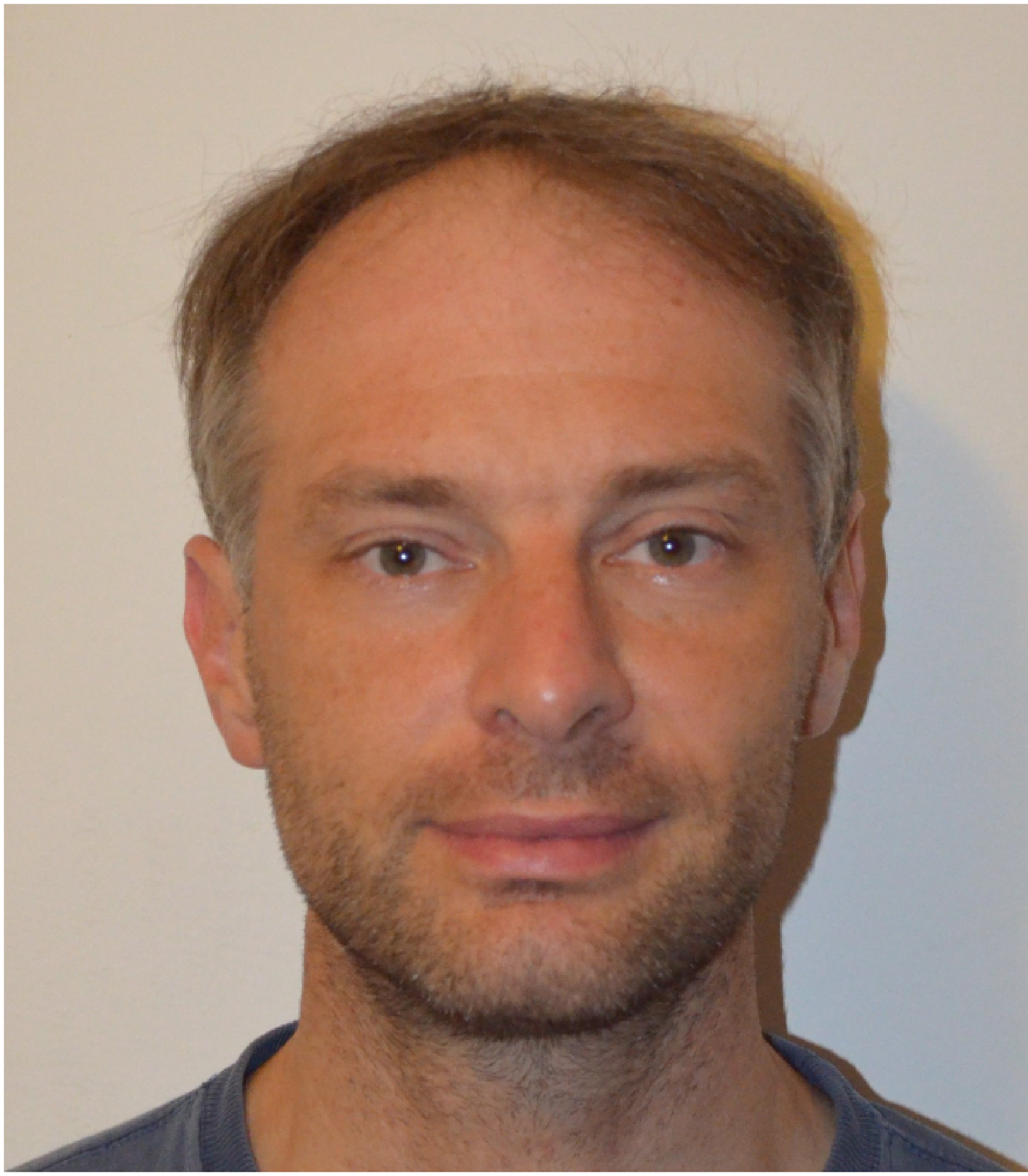}}]
{Alessandro Nordio} (S'00-M'03) is currently a researcher with the Institute of Electronics,
Computer and Telecommunication Engineering of the Italian National Research Council. In 2002 he
received the Ph.D. in Telecommunications from "Ecole Polytechnique Federale de Lausanne", Lausanne,
Switzerland.  From 1999 to 2002, he performed active research with the Department of Mobile
Communications at Eurecom Institute, Sophia Antipolis (France). From 2002 to 2009 he was a post-doc
researcher with the Electronic Department of Politecnico di Torino, Italy.  His research interests
are in the field of signal processing, space-time coding, wireless sensor networks and theory of
random matrices.
\end{IEEEbiography}

\begin{IEEEbiography}[{\includegraphics[width=1in,height=1.25in,clip,keepaspectratio]{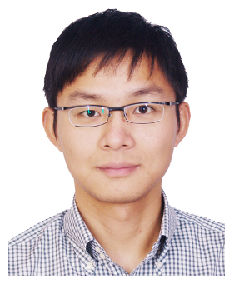}}]
{Siyuan Zhou} (S'14) received the B.S. degree in electronic engineering from Nanjing University of
Posts and Telecommunications, Nanjing, China, in 2008, the M.S. degree in communication and
information systems from Southeast University, Nanjing, China, in 2011. Since Jan. 2012, he has been
pursuing his Ph.D. degree at the Department of Electronics and Telecommunications, Politecnico di
Torino, Torino, Italy. His research lies mainly in the field of random matrix theory applications to
MIMO wireless communication systems.
\end{IEEEbiography}

\end{document}